\documentclass[12pt]{article} 
\usepackage{jheppub}

\usepackage{amsmath,amsfonts,amsbsy,amssymb,array,accents,dsfont}
\usepackage{enumerate}
\usepackage[shortlabels]{enumitem}
\usepackage{graphicx,verbatim} 
\usepackage{caption}
\usepackage{subcaption} 
\usepackage{mathrsfs}
\usepackage{upgreek}
\usepackage{bm}
\usepackage{slashed}
\usepackage{xparse}
\usepackage{xspace}

\NewDocumentCommand\eqn{om}{%
  \IfNoValueTF{#1}
     {\[ #2 \]}
     {\begin{equation}\label{#1} #2  \end{equation} \expandafter\newcommand\csname #1\endcsname{\eqref{#1}\xspace}\ignorespaces}
}
\NewDocumentCommand\eqna{om}{%
  \IfNoValueTF{#1}
    {\begin{align*} #2 \end{align*}}
    {\begin{equation}\label{#1}\begin{split} #2  \end{split}\end{equation} \expandafter\def\csname #1\endcsname{\eqref{#1}\xspace}\ignorespaces}
}

\DeclareFontFamily{U}{mathx}{\hyphenchar\font45}
\DeclareFontShape{U}{mathx}{m}{n}{
      <5> <6> <7> <8> <9> <10>
      <10.95> <12> <14.4> <17.28> <20.74> <24.88>
      mathx10
      }{}
\DeclareSymbolFont{mathx}{U}{mathx}{m}{n}
\DeclareFontSubstitution{U}{mathx}{m}{n}
\DeclareMathAccent{\widecheck}{0}{mathx}{"71}
\DeclareMathAccent{\wideparen}{0}{mathx}{"75}

\newcommand{\rcite}{\cite}

\def\lambdab{{\boldsymbol\lambda}}

\def\zz{{\bf  z}}

\def\sl{\text{sl}}
\def\su{\text{su}}

\def\vareps{\varepsilon}

\def\sltwo{\ensuremath{SL(2,\bR)}}

\def\sutwo{{SU(2)}}

\def\uone{U(1)}

\def\mhat{{\hat\mfm}}
\def\nhat{{\hat\mfn}}

\def\nbarhat{{\hat{\bar{\mfn}}}}

\newcommand{\eq}[1]{\eqref{#1}}

\def\tight#1{\! #1 \!}  % tightens annoying spacing in equations

\def\({\left(}
\def\){\right)}
\def\[{\left[}
\def\]{\right]}

\def\ie{{i.e.}}
\def\eg{{e.g.}}
\def\cf{{c.f.}}
\def\etc{{etc}}

\def\NS{{\sst\rm NS}}

\def\osc{{\rm osc}}

\def\tot{{\rm tot}}

\def\nfive{n_5}

				\def\sfG{{\mathsf G}}		
		\def\sfJ{{\mathsf J}}

		\def\sfL{{\mathsf L}}

\def\sfM{{\mathsf M}}

				\def\sfg{{\mathsf g}}		
				\def\sfk{{\mathsf k}}		
\def\sfm{{\mathsf m}}		\def\sfn{{\mathsf n}}				
						\def\sft{{\mathsf t}}
				\def\sfw{{\mathsf w}}		
\def\sfy{{\mathsf y}}

\def\mfm{{\mathfrak m}}
\def\mfn{{\mathfrak n}}

\DeclareMathSymbol{\medhatsym}{\mathord}{largesymbols}{"62} % basic symbol

\DeclareMathSymbol{\medtildesym}{\mathord}{largesymbols}{"65}% basic symbol
\newcommand\lowermedtildesym{%  adjust height
  \text{\smash{\raisebox{-1.2ex}{%
    $\medtildesym$}}}}
\newcommand\medtilde[1]{% command to be used
  \mathchoice
    {\accentset{\displaystyle\lowermedtildesym}{#1}}
    {\accentset{\textstyle\lowermedtildesym}{#1}}
    {\accentset{\scriptstyle\lowermedtildesym}{#1}}
    {\accentset{\scriptscriptstyle\lowermedtildesym}{#1}}
}
\newcommand{\rhoo}{\ensuremath{\! \rho \:\! }}

\def\half{\frac12}
\def\coeff#1#2{{\textstyle \frac{#1}{#2}}}

\def\tr{{\rm Tr}}
\def\One{{\hbox{1\kern-1mm l}}}

\def\ch{{\rm ch}}

\def\barray{\begin{array}}
\def\earray{\end{array}}
\def\be{\begin{equation}}
\def\ee{\end{equation}}
\def\bea{\begin{align}}
\def\eea{\end{align}}
\def\nn{\nonumber}

\newcommand{\bR}{{\mathbb R}}
\newcommand{\bS}{{\mathbb S}}
\newcommand{\bT}{{\mathbb T}}

\newcommand{\bZ}{{\mathbb Z}}

\definecolor{cardinal}{rgb}{0.6,0,0}
\definecolor{darkgreen}{rgb}{0,0.4,0}
\definecolor{green}{rgb}{0,0.4,0}
\definecolor{golden}{rgb}{0.92, 0.7, 0}
\definecolor{midnight}{rgb}{0, 0, 0.5}
\definecolor{darkblue}{rgb}{0, 0, 0.7}

\numberwithin{equation}{section}

\mathchardef\mhyphen="2D

 \def\cE{\mathcal {E}} 
\def\cG{\mathcal {G}} \def\cH{\mathcal {H}} 
\def\cJ{\mathcal {J}}  \def\cL{\mathcal {L}}
\def\cM{\mathcal {M}} \def\cN{\mathcal {N}} 
 \def\cQ{\mathcal {Q}} \def\cR{\mathcal {R}}
\def\cS{\mathcal {S}}  
\def\cV{\mathcal {V}} \def\cW{\mathcal {W}} 
\def\cY{\mathcal {Y}} \def\cZ{\mathcal {Z}}

\def\k{\sfk}

\def\one{{\hbox{\kern+.5mm 1\kern-.8mm l}}}
\def\zero{{\hbox{0\kern-1.5mm 0}}}

\newcommand{\ket}[1]{{\,| {#1} \rangle}}

\newcommand{\floor}[1]{\left\lfloor #1 \right\rfloor}
\newcommand{\ceil}[1]{\left\lceil #1 \right\rceil}

\newcommand{\T}[3]{\ensuremath{ #1{}^{#2}_{\phantom{#2} \! #3}}}		%general tensor with upper indices displayed first

\def\id{\textrm{id}}

\definecolor{darkviolet}{rgb}{0.58, 0.0, 0.83}

\def\id{{1 \kern-.28em {\rm l}}}

\def\journal#1&#2(#3){\unskip, \sl #1\ \bf #2 \rm(19#3) }
\def\andjournal#1&#2(#3){\sl #1~\bf #2 \rm (19#3) }

\def\ie{{\it i.e.}}
\def\eg{{\it e.g.}}
\def\cf{{\it c.f.}}

\def\etc{{\it etc}}

\def\sst{\scriptscriptstyle}

\def\coeff#1#2{{\textstyle{\frac{#1}{ #2}}}}
\def\half{\frac12}

\def\One{{1\hskip -3pt {\rm l}}}

\catcode`\@=11
\def\slash#1{\mathord{\mathpalette\c@ncel{#1}}}
\def\vareps{\varepsilon}
\def\underrel#1\over#2{\mathrel{\mathop{\kern\z@#1}\limits_{#2}}}

\catcode`\@=12

\def\ket#1{\left| #1\right\rangle}

\def\tr{{\rm tr}}

\def\exp{{\rm exp}}

\def\ch{{\rm cosh}}

\newcommand{\msl}{m}
\newcommand{\msu}{m'}
\newcommand{\Msl}{M}
\newcommand{\Msu}{M'}
\newcommand{\wsl}{w}
\newcommand{\wsu}{w'}

\newcommand{\bmsl}{\bar{m}}
\newcommand{\bmsu}{\bar{m}'}
\newcommand{\bMsl}{\bar{M}}
\newcommand{\bMsu}{\bar{M}'}
\newcommand{\bwsl}{w}
\newcommand{\bwsu}{\bar{w}'}
\newcommand{\jsl}{j}
\newcommand{\jsu}{j'}

\newcommand{\dbydt}{\ensuremath{{\partial}/{\partial t}}}

\def\ie{{\it i.e.}}
\def\eg{{\it e.g.}}

\def\mbar{{\bar m}}

\title{%\LARGE
Cogito, ergo - strings: \\
\Large
Supersymmetric ergoregions and their stringy excitations
}

\author{Emil J. Martinec$^a$, Stefano Massai$^{b,c}$ {\it and}\,
  David Turton$^d$\\}

\affiliation[a]{
Kadanoff Center for Theoretical Physics and Enrico Fermi Institute\\ 
University of Chicago\\ 
5640 S. Ellis Ave.\\
Chicago IL 60637\\ 
}

\affiliation[b]{
Dipartimento di Fisica e Astronomia ``Galileo Galilei''\\
Universit\`a di Padova, Via Marzolo 8, 35131 Padova, Italy\\
}

\affiliation[c]{
INFN, Sezione di Padova, Via Marzolo 8, 35131 Padova, Italy\\
}

\affiliation[d]{
Mathematical Sciences and STAG Research Centre, University of Southampton, \\
Highfield, Southampton, SO17 1BJ, UK\\
}

 \emailAdd{e-martinec@uchicago.edu}
 \emailAdd{stefano.massai@pd.infn.it}
 \emailAdd{d.j.turton@soton.ac.uk}

\abstract{ 
Supergravity solutions describing supersymmetric rotating bound states of NS fivebranes, fundamental strings and momentum can sometimes have an ergoregion but no horizon.
In such supersymmetric ergoregions, there is an unusual feature that there exist BPS ``excitations'' that cost negative (or zero) energy as measured from asymptotic infinity.
We study the spectrum of supergravity and string excitations of these backgrounds using their worldsheet description as gauged Wess-Zumino-Witten (WZW) models, and construct their holographic map to transitions in the dual spacetime CFT.
The backgrounds generically contain orbifold singularities, and we exhibit vertex operators corresponding to twisted sector ground states localized at the orbifold fixed points.
The gauged WZW model thus provides a valuable tool to explore the stringy structure of such heavy BPS states in $AdS_3/CFT_2$.
}

\begin{document}
\hypersetup{pageanchor=false}
\begin{titlepage}
\maketitle
\thispagestyle{empty}
\end{titlepage}
\hypersetup{pageanchor=true}
\pagenumbering{arabic}

%\toc
%\thispagestyle{empty}

%%%%%%%%%%%%%%%%%%%%%%%%%%%%%%%%%%%%%%%%%%%%%%%%%%
%%%%%%%%%%%%%%%%%%%%%%%%%%%%%%%%%%%%%%%%%%%%%%%%%%

\section{Introduction}

A fertile testing ground for gauge/gravity duality is provided by certain geometries sourced by fivebranes carrying either fundamental string or momentum charge, known as supertubes. Two-charge supertube solutions have been completely enumerated, and the correspondence between individual microstate geometries and coherent states in the non-gravitational dual is well understood~\rcite{Lunin:2001fv,Lunin:2001jy,Rychkov:2005ji,Kanitscheider:2007wq,Giusto:2015dfa}.  A particularly well-studied set of examples are the round onebrane-fivebrane supertubes, whose geometry in the AdS decoupling limit is rotating $(AdS_3\times\bS^3)/\bZ_k$~\rcite{Balasubramanian:2000rt,Maldacena:2000dr}; these geometries are dual to particular supersymmetric ground states of the spacetime CFT.  Coherent (spectral flow) excitations of these backgrounds excite a third (momentum) charge, as well as $\bS^3$ angular momentum, while still taking the form of rotating $(AdS_3\times\bS^3)/\bZ_k$. The specific dual states in the spacetime CFT are again known, and consist of Fermi sea excitations above two-charge CFT ground states~\rcite{Lunin:2004uu,Giusto:2004id,Giusto:2004ip,Jejjala:2005yu,Giusto:2012yz,Chakrabarty:2015foa}.

Depending on the choice of spectral flow transformation, these backgrounds can preserve some supersymmetries~\rcite{Lunin:2004uu,Giusto:2004id,Giusto:2004ip,Jejjala:2005yu,Giusto:2012yz} or no supersymmetry~\rcite{Jejjala:2005yu,Chakrabarty:2015foa}.  The most general description of these families of states involves fractional spectral flow, as worked out for the supersymmetric states in~\rcite{Giusto:2012yz}, and for the non-supersymmetric states in~\rcite{Chakrabarty:2015foa}.

As an added bonus, the bulk description of these states (and small excitations around them) is understood not only at the level of supergravity~-- an exactly solvable worldsheet string theory is available~\rcite{Martinec:2017ztd}.
The null-gauged Wess-Zumino-Witten (WZW) model
\be
\label{cosets}
\frac\cG\cH \,=\, \frac{\sltwo\times\sutwo\times \bR_t\times\bS^1_y}{U(1)_L\times U(1)_R} \times \cM ~,
\ee
with $\cH$ a pair of null isometries of $\cG$ and $\cM=\bT^4$ or $K3$, is an exactly solvable 2d CFT which describes all of these round supertube geometries, by varying the choice of embedding $\cH\subset\cG$~\rcite{Martinec:2017ztd,Bufalini:2021ndn}.\footnote{Globally, we work with the universal cover of $\sltwo$ and we gauge a null cylinder, $\mathbb{R}\times S^1$.}

These backgrounds have yielded a variety of insights into gauge/gravity duality, see \eg~the review~\rcite{Skenderis:2008qn}, as well as \eg~the more recent works~\rcite{Galliani:2016cai,Chakrabarty:2021sff}.  They provide a rare example where we can probe the duality map beyond the level of supergravity, to include stringy effects at the non-perturbative level in $\alpha'$~\rcite{Martinec:2018nco,Martinec:2019wzw,Martinec:2020gkv,Bufalini:2022wyp,Bufalini:2022wzu,Martinec:2022okx}.

In this work, we focus on the supersymmetric three-charge spectral flowed supertubes~\rcite{Lunin:2004uu,Giusto:2004id,Giusto:2004ip,Jejjala:2005yu,Giusto:2012yz}, and on a particular known feature~-- that they contain ergoregions with respect to the canonical asymptotically timelike Killing vector field, but no horizons, as we review in Section~\ref{sec:review}.  The ergoregion surrounds the underlying fivebrane source, out to a distance of order the $AdS_3$ radius.
Having the exact perturbative string theory under control, we can study the full spectrum of supergravity and stringy modes that probe the ergoregion, as well as their effect on the background when they are excited.

These supersymmetric ergoregions contain modes that are BPS, but which lower the energy of the background (or keep the energy unchanged). This can happen because these modes reduce the $y$-momentum of the background (as well as the S$^3$ angular momenta). The existence of such modes is implied by the analysis of~\rcite{Giusto:2012yz}, and their backreaction was studied in~\rcite{Chakrabarty:2021sff}. 

Being supersymmetric, these backgrounds are expected to be linearly stable. However, it has been suggested that, classically, they may be non-linearly unstable~\rcite{Eperon:2016cdd}, see also~\rcite{Marolf:2016nwu}. This feature was subsequently argued to correspond not to a runaway process, but instead to slow scrambling motion on a moduli space of nearby states, in which the accumulation of small deformations modifies the background configuration~\cite{Bena:2018mpb,Martinec:2020gkv}.

The physics of these supersymmetric ergoregions contrasts with what happens in the absence of supersymmetry. 
In the asymptotically flat non-supersymmetric solutions in this family, known as the JMaRT solutions after the authors of~\rcite{Jejjala:2005yu}, certain supergravity modes are linearly unstable to ergoregion emission~\rcite{Friedman:1978ygc,Cardoso:2005gj}. 
This process has been studied holographically~\rcite{Chowdhury:2007jx,Chowdhury:2008bd,Chowdhury:2008uj,Avery:2009tu,Chakrabarty:2015foa}.

In the presentation of these backgrounds as gauged WZW models~\eqref{cosets}, there are two time coordinates on $\cG$, namely $\sft\in\bR_t$ and $\tau\in\sltwo$, and hence two notions of energy.  The null gauging construction fuses these two times together into one physical time coordinate transverse to the null gauge orbits.  
These two time coordinates are conjugate to the two natural notions of energy~-- the natural energy $\vareps$ measured at asymptotic infinity, conjugate to $\sft$, that corresponds to energy in the holographic CFT; and the energy measured in the cap of the geometry in the natural co-rotating frame, which aligns with the energy conjugate to $\tau$ in $\sltwo$.  
These two energies are related by the null gauging constraints on physical states.
This relation is mediated by all the other conserved quantum numbers carried by the state which enter the constraints. 
As a result, some modes have positive ``cap energy'' conjugate to $\tau$, but negative ``asymptotic energy'' $\vareps$. These are examples of ergoregion modes~\cite{Martinec:2018nco}.

Although the analysis of part of the spectrum in~\rcite{Giusto:2012yz,Martinec:2018nco} implies the existence of modes with negative asymptotic energy, so far there has been no detailed analysis of the supergravity or stringy spectrum. 
In this work, we provide such an analysis, by studying several aspects of the ergoregions of these supersymmetric backgrounds:
\begin{enumerate}
\item
We extend the analysis of massless ergoregion modes to the entire supergravity spectrum, by analyzing the physical state constraints of the gauged WZW model.  In addition, we map this spectrum to excitations around the corresponding state in the dual spacetime CFT. 
\item
We relate the quantum numbers of string states to the semiclassical support of their wavefunctions (and as a consistency check, demonstrate that modes with negative energy as measured from infinity are predominantly supported in the ergoregion).
\item
We exhibit stringy ergoregion modes, which we refer to as ``ergo-strings'', including twisted sector strings pinned to orbifold singularities in the cap of the geometry.
\end{enumerate}

In Section~\ref{sec:review} we review the backgrounds we study, and their ergoregions. In Section \ref{sec:semiclassics} we analyze null geodesics, which were important in the analysis of~\rcite{Eperon:2016cdd} (see also~\rcite{Chervonyi:2013eja,Bianchi:2017sds}). We work in the AdS decoupling limit, and relate such null geodesics to the properties of harmonics on the group manifold. In particular, we show how the conserved momenta of null trajectories determine their spacetime location, and determine that the trajectories of negative energy sit in the ergoregion.

The exact worldsheet presentation of these backgrounds is reviewed in Section~\ref{sec:sugra-spectrum}, including how the orbifold structure arises as a residual discrete subgroup of the gauge group in a particular gauge choice.  We solve the physical state constraints at the massless level, and exhibit the ergoregion modes for the full range of supergraviton polarizations.

In Section~\ref{sec:stringspec}, we extend the analysis to general string modes, and illustrate the the solution of the constraints with several examples.  First, we extend the supergravity modes in a natural way to sectors of nonzero winding along $\bS^1_y$; second, we analyze orbifold twisted sector ground states, whose condensation resolves the orbifold singularity; lastly, we exhibit ergoregion modes among a simple class of massive string states. 
We conclude our analysis with a discussion in Section~\ref{sec:disc}.  An Appendix reviews details of the limit(s) that decouple the flat spacetime region from the fivebrane throat.

%%%%%%%%%%%%%%%%%%%%%%%%%%%%%%%%%%%%%%%%%%%%%%%%%%
%%%%%%%%%%%%%%%%%%%%%%%%%%%%%%%%%%%%%%%%%%%%%%%%%%
\section{Geometry} 
\label{sec:review}

%%%%%%%%%%%%%%%%%%%%%%%%%%%%%%%%%%%%%%%%%%
\subsection{Three-charge supersymmetric spectral flowed supertubes}
\label{sec:GLMT}

The three-charge supersymmetric spectral flowed supertube backgrounds~\rcite{Lunin:2004uu,Giusto:2004id,Giusto:2004ip,Jejjala:2005yu,Giusto:2012yz} were originally constructed with $\mathbb{R}^{1,4}\times \mathbb{S}^1_y$ asymptotics. In six dimensions, supersymmetry guarantees a globally null Killing vector field~\rcite{Gutowski:2003rg}, which we denote by ${\partial}/{\partial u}$. The supersymmetric spectral flowed supertube backgrounds preserve the supersymmetries that anticommute to translations generated by this null Killing vector.

We shall study these solutions in the fivebrane decoupling limit, as discussed for instance in~\cite{Martinec:2018nco}. For the convenience of the reader, we summarize the steps involved in taking this limit, in the notation and conventions of this work, in Appendix~\ref{sec:asymflat}.
We set $\ell_s=1$, we denote the integer numbers of NS5 branes by $n_5$, and we denote by $n_1$ the integer number of fundamental strings in the background, which is taken to be parametrically large (see Appendix~\ref{sec:asymflat}).
We define
\be
u = t-y \,, \qquad v = t+y \,.
\ee

In the fivebrane decoupling limit, the supergravity solutions take the form
\begin{align}
      ds^2  \,  =&   \,\;  n_5(d\theta^2 + d\rho^2) + \frac{1}{\Sigma}\Bigg[  - f_0\;\! du dv +   \frac{s(s+1) }{k R_y} \;\! \Delta\;\! dv^2  \nonumber \\[1mm]
    & + \Big( n_5 \sinh^2 \rhoo + n_5 (s+1)^2 + \sfk^2 R_y^2 \Big)
    \:\! \sin^2 \theta \;\! d\phi^2 + \Big( n_5 \sinh^2 \rhoo + n_5 s^2 + \sfk^2 R_y^2 \Big) \;\! \cos^2 \theta \:\! d\psi^2  \nonumber \\[1mm]
    & + 2  \left( \!\!\; \Big(   \sfk R_y-(s+1)\Delta \Big)  dy -(s+1) \:\! \Delta \:\! dt \right)  \sin^2 \theta \:\! d\phi + 2 \left( \!\!\: \Big( \sfk R_y+ s\Delta  \Big)  dy + s  \:\! \Delta \:\! dt  \right)  \cos^2 \theta \;\! d\psi \Bigg] \nonumber \\[0.5mm]
  &  + d\zz\cdot d\zz , 
  \label{GLMTmetric2}\\[3mm]
 B  \:  =&   \,\; \frac{1}{\Sigma} \Bigg[  \frac{\sfk R_y}{n_5} \;\! \Delta \:\! dt \wedge dy +  \left( n_5 \sinh^2\rhoo + n_5 \,  (s+1)^2 + \sfk^2 R_y^2  \right) \, \cos^2 \!\theta \,  d\phi \wedge d\psi  
  \nonumber \\[1.5mm]
 & + \left( \!\!\: \Big(  \sfk R_y - (s+1)\Delta\Big) dy - (s+1) \:\!\Delta \:\! dt  \right)\!\!\: \wedge  \cos^2 \theta \:\! d\psi
 +\left( \!\!\: \Big( \sfk R_y + s\Delta \Big)  dy + s \:\! \Delta \:\! dt  \right) \!\!\: \wedge  \sin^2\theta  \:\! d\phi  \Bigg] , \nn \\[3mm]
 e^{2\Phi} \: &
= \frac{\sfk R_y V_4}{n_1} \frac{\Delta}{\Sigma} \;,
\nn
\end{align}
where $\zz$ parametrizes $\bT^4$ whose proper volume is $(2\pi)^4 V_4$, and where
\begin{equation}
\label{GLMTparams}
\Delta =  \sfk R_y + \frac{n_5 s(s+1)}{\sfk R_y} ~,~~~~~
\Sigma = f_0 + \frac{\sfk R_y}{n_5}\Delta
~,~~~~~
f_0 =  \sinh^2\!\rho -s\,\sin^2\!\theta + (s\tight+1)\cos^2\!\theta 
~.
\end{equation}
The parameters $s,\sfk$ are required to satisfy
\be
\frac{s(s+1)}{\sfk} ~\in~\mathbb{Z} \;.
\ee
This condition arises from flux quantization (under the assumption that $\gcd(n_1,n_5)=1$)~\cite{Giusto:2012yz}. It is also necessary for the solutions to have a well-defined holographic interpretation, as this quantity is the quantized momentum per strand of the holographic CFT~\cite{Giusto:2012yz}.

We will also be interested in the AdS$_3$ limit of these solutions. This limit consists of sending $R_y\to\infty$, holding fixed the rescaled energies $ER_y$ and $y$-momenta $P_yR_y$. One can take this limit of the background solutions by defining 
\begin{equation}
\label{eq:ads3-limit}
    \tilde{t} \,=\, \frac{t}{R_y} \,, \qquad
    \tilde{y} \,=\, \frac{y}{R_y} \,,
\end{equation}
and sending $R_y\to \infty$ at fixed $\tilde{t}$, $\tilde{y}$. The resulting solution is asymptotically AdS$_3 \times \mathbb{S}^3 \times \mathbb{T}^4$, and the six-dimensional part of the metric is given by
\begin{align}
\label{eq:glmt-ads}
    \frac{1}{n_5}ds^2 \,=\;\:\! &{}-\frac{1}{\k^2}\cosh^2\rhoo \;\! d\tilde{t}^2 + d\rho^2 +\frac{1}{\k^2}\sinh^2\rhoo \;\! d\tilde{y}^2  \cr
    &{}+d\theta^2 + \cos^2\theta\left( d\psi + \frac{s}{\k} d\tilde{t} + \frac{s+1}{\k}d\tilde{y} \right)^2
    + \sin^2\theta\left( d\phi - \frac{s+1}{\k} d\tilde{t} - \frac{s}{\k}d\tilde{y} \right)^2 .
\end{align}
The following large coordinate transformation maps these decoupled geometries to orbifolds of the non-rotating  global AdS$_3 \times$S$^3$ NS-NS vacuum solution, and is known as (fractional) spacetime spectral flow~\cite{Giusto:2012yz}, see also~\cite{Lunin:2004uu,Giusto:2004id,Giusto:2004ip}:
\begin{align}
\label{eq:sf-GLMT-sugra}
    \psi_{\NS}  \,=\;\:\! & \psi + \frac{s}{\k} \tilde{t} + \frac{s+1}{\k}\tilde{y}  ~,\quad~~
    \phi_{\NS}
    \,=\, \phi - \frac{s+1}{\k} \tilde{t} - \frac{s}{\k} \tilde{y} ~,\quad~~
    \tilde{t}_{\NS} \,=\, t ~,
    \quad~~
    \tilde{y}_{\NS} \,=\, \tilde{y} ~.
\end{align}
For generic $s,\sfk$, the coordinate identification $\tilde{y} \sim \tilde{y} + 2 \pi $ induces orbifold singularities via the identification~\cite{Giusto:2012yz,Jejjala:2005yu}
\begin{align}
\label{eq:sugra-orbact-GLMT}
\big(\tilde{y},\psi_{\NS},\phi_{\NS}\big) 
\; \sim \;
\big(\tilde{y},\psi_{\NS},\phi_{\NS}\big)  +
2\pi \Big(1,\frac{s+1}{\k},-\frac{s}{\k}\Big) \,. 
\end{align}
We will sometimes refer to the general family of these backgrounds as the GLMT backgrounds, after the authors of~\cite{Giusto:2012yz}.

%%%%%%%%%%%%%%%%%%%%%%%%%%%%%%%%%
\subsection{Supersymmetric ergoregions}
\label{sec:ergoregion}

We now discuss the ergoregions of the supersymmetric solutions discussed above. 
First, we recall that when a spacetime has isometries that correspond to spatial directions of finite asymptotic size, there is no preferred definition of an ergoregion. 
For each asymptotically timelike Killing vector field, one can ask whether there is a region of spacetime where it becomes spacelike, in which case there is an ergoregion corresponding to that Killing vector field; see e.g.~\cite{Pelavas:2005}.

In the fivebrane decoupling limit, the $y$ circle and S$^3$ asymptote to fixed proper sizes, so there is indeed a family of asymptotically timelike Killing vector fields. 
However, the asymptotically timelike  Killing vector field $\dbydt$ plays a preferred role, for two reasons. First, the energy measured with respect to $\dbydt$ corresponds to the energy of the holographically dual CFT. Second, in the corresponding asymptotically flat five-dimensional solutions obtained by reducing on the $y$ circle, $\dbydt$ is the unique asymptotically timelike Killing vector field.

We therefore focus on the ergoregion corresponding to $\dbydt$. The norm of $\dbydt$ is given by $g_{tt}$. So the six-dimensional ergosurface is defined by $g_{tt}=0$ and the six-dimensional ergoregion is defined by $g_{tt}>0$.

In the solutions \eqref{GLMTmetric2}, there is an ergoregion with respect to $\dbydt$, as noted  in~\rcite{Jejjala:2005yu}, since
\be
\qquad\qquad
g_{tt} \;=\; 
- \frac{ f_0 - \Delta_p }{\Sigma} \;, 
\qquad\qquad  
\Delta_p \;\equiv\; s(s+1)+\nfive \left(\frac{s(s+1) }{k R_y}\right)^2 \,.
\label{eq:ergoregion-glmt}
\ee
The 6D ergosurface is the locus $f_0=\Delta_p$ and the 6D ergoregion is given by $f_0<\Delta_p$. 
We note that $\Delta_p > 0$ when $s>0$, and that $f_0$ takes its minimum value at $\rho=0$ and $\theta=\pi/2$, where $f_0=-s$.
In the AdS$_3$ limit, $\Delta_p$ has the finite limit $s(s+1)$.

Generic excitations of these supersymmetric backgrounds have positive energy. However, due to the ergoregion, there are also ``excitations'' that {\it reduce} the absolute value of both the energy (measured with respect to ${\partial}/{\partial t}$) and $y$-momentum of the background. Moreover, some of these excitations preserve the BPS relation. A subset of such excitations are present in the analysis of linearized supergraviton excitations in~\cite{Giusto:2012yz}. The backreaction of collective excitations of some of these modes was studied in~\cite{Chakrabarty:2021sff}. We will identify the corresponding massless vertex operators in due course. 

In addition to modes that lower the energy, there are supersymmetric modes that exactly preserve the energy of the background. One way to see this is that there is a family of null geodesics that are everywhere tangent to the Killing vector corresponding to supersymmetry, ${\partial}/{\partial u}$. These geodesics have the property that their energy 
is equal to their $y$-momentum.  
Moreover, when $s>0$, the surface $f_0=0$ is inside the 6D ergoregion, since $\Delta_p>0$.  On this surface, the Killing vector fields ${\partial}/{\partial u}$ and ${\partial}/{\partial y}$ are orthogonal, and such geodesics have $E_6=p_y=0$, as noted in~\cite{Eperon:2016cdd}.

We thus see that there is a rich set of features associated with these supersymmetric ergoregions. We will analyze these features via both classical geodesics and worldsheet vertex operators in this paper.

As an aside, we recall that the surface $f_0=0$ is also important in the asymptotically flat five-dimensional solutions. These solutions do not have ergoregions, however, there is a surface at which the five-dimensional Killing vector field $\dbydt$ becomes null. 
This surface is given by $f_0=0$, and is known as the {\it evanescent ergosurface}~\cite{Gibbons:2013tqa}.
We also note that in the limit $s=0$, the spectral flowed solutions reduce to two-charge circular supertubes~\cite{Balasubramanian:2000rt,Maldacena:2000dr}. These do not have a 6D ergoregion, but rather a 6D evanescent ergosurface at $f_0=0$, on which $\dbydt$ becomes null. Correspondingly, the energy of excitations is bounded below by zero in the two-charge limit.

Due to the globally null Killing vector field, it is expected that the supersymmetric solutions do not suffer from a linear instability.
However, the supersymmetric solutions have long-lived quasi-normal modes~\cite{Eperon:2016cdd}, see also~\cite{Chakrabarty:2019ujg}.
It has been argued that these long-lived modes may give rise to a classical non-linear instability~\cite{Eperon:2016cdd}, see also~\cite{Marolf:2016nwu}; this phenomenon has been argued to be better described by slow scrambling dynamics on the moduli space of nearby states~\cite{Bena:2018mpb,Martinec:2020gkv}.

%%%%%%%%%%%%%%%%%%%%%%%%%%%%%%%%%
\section{Geodesics and Wavefunctions}
\label{sec:semiclassics}

In the semi-classical limit, supergraviton wavefunctions localize along null geodesics.  One can discern many of the features of the wavefunctions from a study of such geodesics.  In this section, we use the group symmetry of the GLMT (and JMaRT) geometries in the $AdS_3$ decoupling limit to study the semiclassical dynamics of supergravitons.  

We discuss two approaches to geodesic motion.  One is group theoretic~-- exponentiating a generator of the group, and labeling the resulting geodesic by its eigenvalues under the currents $J^3,\bar J^3$ and the quadratic Casimir $\vec J^{\,2}$.  The other employs the formalism of Lagrangian and Hamiltonian mechanics of a particle on the group manifold.  

The group manifolds $\sltwo$ and $\sutwo$ play two roles here. 
First, $\sltwo\times\sutwo$ are the $\sfk$-fold covering space of 
the 6d part of the physical 10-dimensional spacetime \eqref{eq:glmt-ads}. In addition, both $\sltwo$ and $\sutwo$ are factors in the ``upstairs'' $10+2$-dimensional group manifold target space of the null gauging construction that we shall describe in Section~\ref{sec:sugra-spectrum}.

%%%%%%%%%%%%%%%%%%%%%%%%%%%%%%%%%
\subsection{Geodesics and classical strings on \texorpdfstring{$\sutwo$}{}}

At high momentum, wavefunctions are concentrated along semi-classical trajectories, which are geodesics on the group manifolds $\sltwo$ and $\sutwo$. 
We begin with $\sutwo$ and write the line element as
\be
ds^2 = d\theta^2 
+ \cos^2\theta \;\!  d\psi^2 
+ \sin^2\theta \;\!  d\phi^2 \,.
\ee
This metric is related to the parametrization of the group in terms of Euler angles
\be
\label{gsumat1}
g_\su \;=\; e^{-i(\phi-\psi)\sigma_3/2}\, e^{i(\pi/2-\theta)\sigma_2}\, e^{-i(\phi+\psi)\sigma_3/2} \;=\; 
\left(\begin{matrix} e^{-i\phi}\sin\theta ~&~  e^{i\psi}\cos\theta \\ - e^{-i\psi}\cos\theta ~&~ e^{i\phi} \sin\theta \end{matrix}\right) ~.
\ee
Classical solutions to the $\sutwo$ WZW model take the form
\be
\label{gclassical}
g(z,\bar z) =  g_\ell(z) g_r(\bar z) ~.
\ee
A simple solution is geodesic motion along the $\phi$ circle at $\theta=\frac\pi2$, which corresponds to the classical solution
\be
\label{suhwtgeod}
g_\ell(z) = e^{-i\nu' z \sigma_3/2}
~~,~~~~
g_r(\bar z) = e^{-i\nu' \bar z \sigma_3/2} ~,
\ee
where $z,\bar z=\xi_0\pm\xi_1$ are worldsheet coordinates.
One can then rotate this geodesic to some other great circle on $\bS^3$ via
\be
\label{sugengeod}
g_\ell(z) =  e^{-i\alpha'_\ell \sigma_1/2}  \,  e^{-i\nu' z \sigma_3/2} 
~~,~~~~
g_r(\bar z) =  e^{-i\nu' \bar z \sigma_3/2}  \,  e^{-i\alpha'_r \sigma_1/2} ~.
\ee
Multiplying out the group elements, one finds the geodesic motion 
\be
\label{geodesic gsu}
g(\xi_0) = 
\left(\begin{matrix} 
\cos{\nu' \xi_0}\cos\frac{\alpha'_\ell+\alpha'_r}2 
-i\sin{\nu' \xi_0}\cos\frac{\alpha'_\ell-\alpha'_r}2 
~~&~  
-i\cos{\nu' \xi_0}\sin\frac{\alpha'_\ell+\alpha'_r}2 
+\sin{\nu' \xi_0}\sin\frac{\alpha'_\ell-\alpha'_r}2
\\[.3cm] 
-i\cos{\nu' \xi_0}\sin\frac{\alpha'_\ell+\alpha'_r}2 
-\sin{\nu' \xi_0}\sin\frac{\alpha'_\ell-\alpha'_r}2 
~~&~~ 
\cos{\nu' \xi_0}\cos\frac{\alpha'_\ell+\alpha'_r}2 
+i\sin{\nu' \xi_0}\cos\frac{\alpha'_\ell-\alpha'_r}2 
\end{matrix}\right) ~.
\ee
%%%%%%%%%%%%%%%%%%%%%%%%%%%%
Comparing this matrix to \eqref{gsumat1}, one finds a trajectory that oscillates in $\theta$ according to
\be
\label{thetamotion}
-\cos 2\theta = 
\cos(\alpha'_\ell\tight+\alpha'_r)\cos^2(\nu'\xi_0) + \cos(\alpha'_\ell\tight-\alpha'_r) \sin^2(\nu'\xi_0)~.
\ee
The maximum and minimum values of the polar coordinate $\theta$ are
\be
\label{thetapm}
\theta_\pm =\half\Big( \pi - \big|\alpha'_\ell\mp\alpha'_r \big|\Big) ~.
\ee

The geodesic motion respects the conservation of the Hamiltonian as well as the Killing momenta associated to left and right motions generated by $J^3$.  The corresponding conserved charges are
\begin{align}
\label{suqnums}
\cE & \,=\, \frac{\nfive}2\tr\bigl[\partial g \;\! \partial g^{-1}\bigr] + \frac{\nfive}2\tr\bigl[\bar\partial g \;\! \bar\partial g^{-1}\bigr] \,=\, \frac{\nfive}2 (\nu')^2 ~,
\nn\\[.2cm]
 J^3_{\su} & \,=\, -\frac{i\nfive}2 \tr\bigl[(\partial g)g^{-1}\sigma_3\bigr] \,=\,
\nfive \left(
\cos^2 \theta \;\! \partial\psi - \sin^2 \theta \;\! \partial\phi
\right)
 \,=\,
 -\frac{\nfive}2\,\nu'\cos(\alpha'_\ell) ~,
\\[.2cm]
\bar J^3_{\su} & \,=\, -\frac{i\nfive}2 \tr\bigl[g^{-1}(\bar\partial g)\sigma_3\bigr] \,=\,
- \nfive \left(
\cos^2 \theta \;\! \bar\partial\psi + \sin^2 \theta \;\! \bar\partial\phi
\right)
 \,=\, -\frac{\nfive}2\,\nu' \cos(\alpha'_r)
  ~~;
\nn
\end{align}
in anticipation of the quantization of the dynamics, these quantities are related to the (half) integer quanta $j',m',\bar m'$ of $\sutwo$ representation theory via
\be
\label{jmsu}
j' \,=\, \frac\nfive2\nu'
~~,~~~~
\frac{\msu}{\jsu} \,=\, -\cos\big(\alpha'_\ell\big)
~~,~~~~
\frac{\bmsu}{\jsu} \,=\, -\cos\big(\alpha'_r\big)  ~.
\ee
The unitary range of allowed values in affine $\sutwo$ representation theory is $0\le\nu'\le1$,%
\footnote{The unitarity bound for quantized strings on the $\sutwo$ and $\sltwo$ group manifolds is seen in the classical theory as a bound on stationary solutions~-- when the momentum exceeds a particular value, the Lorentz force of the background B-field exceeds the string tension and pulls the string apart.  See for instance~\rcite{Martinec:2020gkv} Section 5 for a discussion.} 
while $0\le\alpha'_{\ell,r}\le\pi$ code $m',\bar m'$ (or rather coherent states thereof).%
\footnote{The classical solution corresponds to the limit of large $\nfive$, with $\nu'$ held fixed, and so does not distinguish between $(\nu')^2=(\frac{j'}\nfive)^2$ and $(\nu')^2=\frac{j'(j'+1)}{\nfive^2}\,$.}
In particular, $\alpha'_\ell=0$ corresponds to $m'=-j'$, and similarly $\alpha'_r=0$ corresponds to $\bar m'=-j'$; then the solution~\eqref{suhwtgeod} corresponds to the lowest-weight state.  Holding say $\alpha_r=0$ and dialing $\alpha_\ell$ coherently excites larger values of $m'$, resulting in circular trajectories concentrated at fixed latitude lines $\theta_+=\theta_-=\half(\pi-|\alpha'_\ell|)$.  Dialing both $\alpha'_{\ell,r}$ results in a trajectory that oscillates between $\theta_+$ and $\theta_-$ given by~\eqref{thetapm}, and corresponds to a coherent excitation of both $m',\bar m'$ away from the lowest-weight state.

%%%%%%%%%%%%%%%%%%%%%%%%%%%%%%%%%
\subsection{Geodesics and classical strings on \texorpdfstring{$\sltwo$}{}}
\label{sec:sltwo-geodesics}

Similarly, we consider geodesics on global $AdS_3$. 
We work on the universal cover of $\sltwo$, \ie~with non-compact time direction. We write the line element as
\be
\label{eq:ads-met}
ds^2 \,=\,
{}-\cosh^2\rhoo \;\! d{\tau}^2 + d\rho^2 +\sinh^2\rhoo \;\! d\sigma^2  ~.
\ee
We work with the Euler angle parametrization of $\sltwo$, 
\be
\label{gslmat}
g_\sl \,=\,
e^{i(\tau+\sigma)
\sigma_3/2}\, e^{\rho\sigma_1}\, e^{i(\tau-\sigma)\sigma_3/2} \,=\, 
\left(\begin{matrix} e^{i\tau}\cosh\rho ~&~  e^{i\sigma}\sinh\rho \\ e^{-i\sigma}\sinh\rho ~&~ e^{-i\tau} \cosh\rho \end{matrix}\right) ~.
\ee
Highest weight states in the discrete series representation of $\sltwo$ (see \eg~\cite{Maldacena:2000hw}) correspond to geodesics given by
\be
\label{slhwtgeod}
g_\ell(z) \,=\, e^{i\nu z \sigma_3/2}
~~,~~~~
g_r(\bar z) \,=\, e^{i\nu \bar z \sigma_3/2} ~.
\ee
The matrix $g_\sl$ is diagonal, and so $\rho=0$; the geodesic sits at the center of $AdS_3$, and runs up the time axis at a velocity $\nu$.

The boost transformation
\be
\label{slgengeod}
g_\ell(z) \,=\,  e^{\alpha_\ell \sigma_1/2}  \,  e^{i\nu z \sigma_3/2} 
~~,~~~~
g_r(\bar z) \,=\,  e^{i\nu \bar z \sigma_3/2}  \,  e^{\alpha_r \sigma_1/2} 
\ee
leads to a geodesic trajectory
\be
\label{geodesic gsl}
g(\xi_0) \,=\, 
\left(\begin{matrix} 
\cos{\nu\xi_0}\cosh\frac{\alpha_\ell+\alpha_r}2 
+i\sin{\nu\xi_0}\cosh\frac{\alpha_\ell-\alpha_r}2 
~~&~  
\cos{\nu\xi_0}\sinh\frac{\alpha_\ell+\alpha_r}2 
-i\sin{\nu\xi_0}\sinh\frac{\alpha_\ell-\alpha_r}2
\\[.3cm] 
\cos{\nu\xi_0}\sinh\frac{\alpha_\ell+\alpha_r}2 
+i\sin{\nu\xi_0}\sinh\frac{\alpha_\ell-\alpha_r}2 
~~&~~ 
\cos{\nu\xi_0}\cosh\frac{\alpha_\ell+\alpha_r}2 
-i\sin{\nu\xi_0}\cosh\frac{\alpha_\ell-\alpha_r}2 
\end{matrix}\right) ~,
\ee
which oscillates radially according to
\be
\cosh 2\rho \,=\, \cos^2(\nu\xi_0)\,\cosh(\alpha_\ell\tight+\alpha_r) + \sin^2(\nu\xi_0)\,\cosh(\alpha_\ell\tight-\alpha_r) ~,
\ee
and thus orbits the center of $AdS_3$ between
\be
\label{rhopm}
\rho_\pm \,=\, \half\,\big|\alpha_\ell\tight\pm \alpha_r\big| ~.
\ee
Similarly, one has the $\sltwo$ conserved quantum numbers 
\begin{align}
\label{slqnums}
\cE &= -\frac{\nfive}2\tr\bigl[\partial g \;\! \partial g^{-1}\bigr] - \frac{\nfive}2\tr\bigl[\bar\partial g \;\! \bar\partial g^{-1}\bigr] \,=\, -\frac{\nfive}2 \, \nu^2 ~,
\nn\\[.2cm]
J^3_{\sl} & \,=\, -\frac{i\nfive}2 \tr\bigl[(\partial g)g^{-1}\sigma_3\bigr] 
\,=\, {\nfive}\big( \cosh^2\!\rho\,\partial\tau-\sinh^2\!\rho\,\partial\sigma\big)
\,=\, \frac{\nfive}{2}\nu\cosh(\alpha_\ell)  ~,
\\[.2cm]
\bar J^3_{\sl} & \,=\, -\frac{i\nfive}2 \tr\bigl[g^{-1}(\bar\partial g)\sigma_3\bigr] 
\,=\, {\nfive}\big( \cosh^2\!\rho\,\bar\partial\tau+\sinh^2\!\rho\,\bar\partial\sigma\big)
\,=\, \frac{\nfive}{2}\,\nu \cosh(\alpha_r)
~~; 
\nn
\end{align}
again, in terms of the labels $(j,m,\bar m)$ of $\sltwo$ representation theory, we have 
\be 
\label{eq:m-alpha-l}
j=\frac{\nfive}2\nu
~~,~~~~
\frac\msl\jsl \,=\, \cosh\alpha_\ell
~~,~~~~
\frac\bmsl\jsl \,=\, \cosh\alpha_r
~.
\ee
The lowest weight state $m=\bar m=j$ has $\alpha_\ell=\alpha_r=0$.  Increasing $\alpha_\ell$ keeping $\alpha_r=0$ increases~$m$ while maintaining $\bar m=j$, resulting in a circular orbit at radius $\rho_*=\rho_\pm=\alpha_\ell$.  Increasing~$\bar m$, the orbit spreads between minimum and maximum values $\rho_\pm$ given by~\eqref{rhopm}, until at $m=\bar m$ the motion is purely radial.  Increasing $\bar m$ further, once again the orbit becomes elliptical between the radii $\rho_\pm$.

%%%%%%%%%%%%%%%%%%%%%%%%
%%%%%%%%%%%%%%%%%%%%%%%%

\subsection{Geodesics in spectral flowed supertube backgrounds}

We now discuss geodesics on the GLMT backgrounds in the AdS$_3$ decoupling limit, \eqref{eq:glmt-ads}. 
Since most of our discussion in this subsection involves geodesics on the $\sltwo\times\sutwo$ group manifold, we will carry out the analysis with a slightly more general parametrization of the spacetime spectral flow than~\eqref{eq:sf-GLMT-sugra}, that also includes the non-supersymmetric JMaRT backgrounds~\rcite{Jejjala:2005yu,Chakrabarty:2015foa}. These backgrounds are parametrized by two independent integers $\sfm,\sfn$, with
\be
\label{JMaRT params}
\sfm \,=\, s+1 \,,
\qquad
\sfn \,=\, s \,,
\ee
being the specialization to the supersymmetric spacetimes~\eqref{eq:glmt-ads}.  We will temporarily use this more general parametrization.

Geodesic motion of a particle of mass $\sfM$ is governed by the action
\be
\label{Sgeod}
\cS = \int \!d\xi\sqrt{\sfg}\Big(\sfg^{\xi\xi} \partial_\xi X^\mu\partial_\xi X^\nu G_{\mu\nu}(X) 
 +\sfM^2\Big) ~.
\ee
Variation of this action w.r.t.~$\sfg$ imposes the Hamiltonian constraint 
\be
\label{Hamcon}
\sfg^{\xi\xi} \partial_\xi X^\mu\partial_\xi X^\nu G_{\mu\nu}(X) = \sfM^2 ~.
\ee
We choose the gauge $\sfg=1$ and write $\dot{X}^\mu \equiv \partial_\xi X^\mu$.

We consider the line element \eqref{eq:glmt-ads}, in the $\sfk$-fold covering space coordinates $\tau = \tilde t/\sfk $, $\sigma=\tilde y/\sfk $. 
With the above gauge choice, geodesics are described by the Lagrangian 
\begin{align}
\label{groupL}
\cL_{g} \;=\; \cL_\sl+\cL_\su =
\frac{\nfive}{2} &\Big[\Big( 
-\cosh^2\rhoo \, \dot{\tau}^2 + \dot{\rho}^2 + \sinh^2\rhoo \, \dot{\sigma}^2 \Big)
+\Big(\dot{\theta}^2 + \cos^2\theta \, \dot\psi_{\NS}^2 
+ \sin^2\theta \,\dot\phi_{\NS}^2
\Big)\Big] ~,
\end{align}
where from Eq.\;\eqref{eq:sf-GLMT-sugra} with the replacements $s+1 \to \sfm$, $s \to \sfn$, we have \begin{align}
\label{NS2JMaRT}
    \dot{\psi}_{\NS} \,&=\, 
    \dot\psi + \sfn \;\! \dot{\tau}+ \sfm \;\! \dot{\sigma} ~~,
    \qquad\quad
    \dot{\phi}_{\NS} \,=\, 
    \dot\phi - \sfm \;\! \dot{\tau} - \sfn \;\! \dot{\sigma} ~~.
\end{align}

The isometries corresponding to the Killing vector fields $\partial_\tau$, $\partial_\sigma$, $\partial_\psi$, $\partial_\phi$ commute, and thus give rise to conserved momenta that Poisson commute, leading to the constants of motion
\begin{align}
\begin{aligned}
\label{cons mom}
-p_\tau \;&=\; \nfive\big(\cosh^2\rhoo \, \dot{\tau} - \sfn \cos^2\theta \;\! \dot{\psi}_{\NS}
+ \sfm \sin^2\theta \;\! \dot{\phi}_{\NS}\big) \;,\\
p_\sigma \;&=\; \nfive\big(\sinh^2\rhoo \, \dot{\sigma} + \sfm \cos^2\theta \;\! \dot{\psi}_{\NS}
- \sfn \sin^2\theta \;\! \dot{\phi}_{\NS}\big) \;,\\
m_\psi \;&=\; \nfive\big(\cos^2\theta \;\! \dot{\psi}_{\NS}\big) \;,\\
m_\phi \;&=\; \nfive\big(\sin^2\theta \;\! \dot{\phi}_{\NS}\big) \,.
\end{aligned}
\end{align}
Substituting the second pair into the first pair, 
and introducing the constants $\vareps= -p_\tau/\sfk$ and $n_y = p_\sigma/\sfk$,
we obtain
\begin{align}
\begin{aligned}
\label{eq:geod-sf}
\sfk \vareps \;&=\; \nfive\cosh^2\rhoo \, \dot{\tau} - \sfn \;\! m_\psi
+ \sfm \;\! m_\phi \;,\\
\sfk \:\! n_y 
\;&=\; \nfive\sinh^2\rhoo \, \dot{\sigma} + \sfm \;\! m_\psi
- \sfn \;\! m_\phi  \;,
\end{aligned}
\end{align}
implying that $\cosh^2\rhoo \, \dot{t}$ and $\sinh^2\rhoo \, \dot{y}$ are constants of motion.

In terms of the quantum numbers $\msl,\bmsl,\msu,\bmsu$ introduced in the previous subsections, 
we identify 
\begin{align}
\begin{aligned}
\label{eq:dots-m-mb}
\nfive\cosh^2\rhoo \, \dot{\tau} \;=\; \msl+\bmsl
\quad&,\qquad
m_\phi =
\msu +\bmsu ~,
\\
-\nfive\sinh^2\rhoo \, \dot{\sigma} \;=\; \msl-\bmsl
\quad&,\qquad
m_\psi =
\bmsu - \msu ~.
\end{aligned}
\end{align}
Combining these with the previous set of equations, and making the substitutions~\eqref{JMaRT params}, we obtain
\begin{align}
\begin{aligned}
\label{eq:geod-sf-2}
\sfk \vareps \;&=\; \msl+\bmsl + (2s+1) \msu + 
\bmsu\;,\\
-\sfk n_y &\,=\, \msl- \bmsl + (2s+1) \msu - 
\bmsu  \, . 
\end{aligned}
\end{align}
In order to connect with the discussion in Section~\ref{sec:ergoregion}, we note that $\vareps$ corresponds to the energy measured with respect to $\partial/\partial \tilde{t}$ of the rotating spacetimes \eqref{eq:glmt-ads}.
By contrast, $m+\bar{m}$ is the energy measured with respect to the different Killing vector $\partial/\partial \tilde{t}_{\NS}$ of the non-rotating global AdS$_3 \times$S$^3$ vacuum solution, see \eqref{eq:sf-GLMT-sugra}. 
These equations give the relation between these two energies (and similarly for the two  different AdS$_3$ angular momenta, $n_y$ and $\bmsl-\msl$).
Deep inside the AdS$_3$ region, $\msl+\bmsl$ is a natural local measure of energy, being in the natural co-rotating frame. 
Due to the rotation in the backgrounds, and the presence of an ergoregion, it is possible to have $\msl+\bmsl >0$, while $\varepsilon <0 $. In due course, we shall see that this will be an important feature of certain ergoregion modes. The equations \eqref{eq:geod-sf-2} will reappear in the worldsheet analysis, in Eq.~\eqref{va-glmt-null}.

To completely integrate the equations of motion, one needs two more conserved quantities that Poisson commute with the four momenta in Eq.~\eqref{cons mom}.  These are provided by the individual Hamiltonians $\cH_\sl=\cL_\sl$, $\cH_\su=\cL_\su$ for geodesic motion on $\sltwo$ and $\sutwo$, which in group theoretic terms are the quadratic Casimirs given by the kinetic terms in Eq.~\eqref{groupL}.

Expressing these Hamiltonians in terms of the conserved momenta, we obtain reduced Hamiltonians for the radial and polar motions,
\begin{align}
\label{Heffs}
\begin{aligned}
\cH_\sl &\;=\; n_5^2\, \dot\rho^2
+\frac{(\msl-\bmsl)^2}{\sinh^2\rhoo}
 -\frac{(\msl+\bmsl)^2}{\cosh^2\rhoo}
 \;=\; -4j^2 \,, \\[.2cm]
\cH_\su &\;=\; n_5^2\,\dot\theta^2
+\frac{(\msu-\bmsu)^2}{\cos^2\theta}
+\frac{(\msu+\bmsu)^2}{\sin^2\theta}
\;=\; +4(j')^2 \,,
\end{aligned}
\end{align}
where the normalization of the constants on the right-hand side has been fixed by consistency with the previous subsections.
The Hamiltonian constraint \eqref{Hamcon} is
\be
\cH_\tot\,=\,\cH_\sl+\cH_\su\,=\,\sfM^2 ~~.
\ee
So, for massless particles, we have $j=j'$.

From these relations, setting $\dot\theta$ and $\dot\rho$ to zero, one recovers the turning points of the classical motion~\eqref{thetapm}, \eqref{rhopm}.  Finally, we obtain the trajectories in the GLMT (and JMaRT) geometries from those on the factorized group manifold via the coordinate shifts~\eqref{NS2JMaRT}, followed by a $\bZ_k$ orbifold quotient.

For the GLMT backgrounds, of particular interest are the BPS null geodesics.  These have $\bmsl=j$, $\bmsu=-j$, or in other words $\alpha_r=\alpha'_r=0$.  Thus $\rho_+=\rho_-$ and $\theta_+=\theta_-$, and the motion is stationary in both radial and polar directions (in fact, the equations of motion imply that $\dot\phi=\dot\psi=0$ as well). 
It is useful to write the left-moving momenta $\msu$, $\msl$ in terms of their lowest-weight values (in $D_j^+$ for $\sltwo$) and offsets as
\be
\label{eq:mhat-nhat-defn}
\msu=-j+\mhat ~, 
\qquad
\msl=j+\nhat ~.
\ee
Then the offsets $\mhat$, $\nhat$ determine the location of the BPS null geodesic to be given by
\be
\label{location}
\sinh^2\rho = \frac{\nhat}{2j}
~~,~~~~
\cos^2\theta = \frac{\mhat}{2j} ~.
\ee
One can extend the above analysis to the NS5-brane decoupling limit (\ie\ finite $R_y$) version of these solutions.

%%%%%%%%%%%%%%%%%%%%%%%%%%%%%%%%%
\subsection{Wavefunctions}
\label{sec:wavefns}

The eigenfunctions of the scalar Laplacian on the $\sutwo$ group manifold (Wigner functions) reflect the above semi-classical features.  These (unnormalized) wavefunctions are
\begin{align}
\label{su2wavefns}
D_{j'm'\bar m'}(\theta,\phi,\psi) &= e^{-im'(\phi+\psi)}\,e^{-i\bar m'(\phi-\psi)}\, d_{j'm'\bar m'}(\theta) ~,
\nn\\[.2cm]
d_{j'm'\bar m'}(\theta) &= 
(\cos\theta)^a (\sin\theta)^b P_{q}^{(a,b)}(\theta) ~,
\\[.2cm]
P_{q}^{(a,b)}(\theta) &= \sum_{p=0}^q \bigg(\begin{matrix}{q+a}\\
{q-p}\end{matrix}\bigg)
\bigg(\begin{matrix}{q+b}\\{p}
\end{matrix}
\bigg) \big(
\sin\theta\big)^{2p}
\big(\cos\theta\big)^{2q-2p}~,
\nn
\end{align}
where $a=|\msu-\bmsu|$, $b=|\msu+\bmsu|$, $q=j'-\mu$, with $\mu=\max(|\msu|,|\bmsu|)$.

For $\bar m'=-j'$ ($\alpha'_r=0$), the sum over $p$ collapses to a single (constant) term; the trigonometric polynomial $d_{j'm'\bar m'}(\theta)$ has a single peak at $\theta_*=\theta_+=\theta_-$ determined by the value of $m'$ given by $\alpha'_{\ell}\,$ via~\eqref{thetapm}, \eqref{suqnums}.  
As one increases $\bar m'$ starting from $\bar m'=-j'$, the wavefunction gains $j'+\bar m'$ nodes and has support between $\theta_-$ and $\theta_+$, vanishing rapidly outside this range (over a width $\delta\theta\sim(j')^{-1/2}$), until $\bar m'=-|m'|$, at which point either $\theta_+=\frac\pi2$ (for $m'<0$) or $\theta_-=0$ (for $m'>0$).  At this point, the wavefunction stops gaining nodes; rather its center migrates (toward lower $\theta$ until $\theta_-=0$, for $m'<0$; or toward larger $\theta$, until $\theta_+=\frac\pi2$, for $m'>0$).  At this point, the number of nodes starts decreasing until at $\bar m'=j'$, the wavefunction once again has no nodes, and is now centered at $\theta=\frac\pi2-\theta_*$.  At large $j'$, these are just the WKB wavefunctions one would expect for the classical trajectories~\eqref{sugengeod}, \eqref{thetamotion}. 

\begin{figure}[ht]
\centering
  \begin{subfigure}[b]{0.38\textwidth}
  \hskip 0cm
\includegraphics[width=\textwidth]{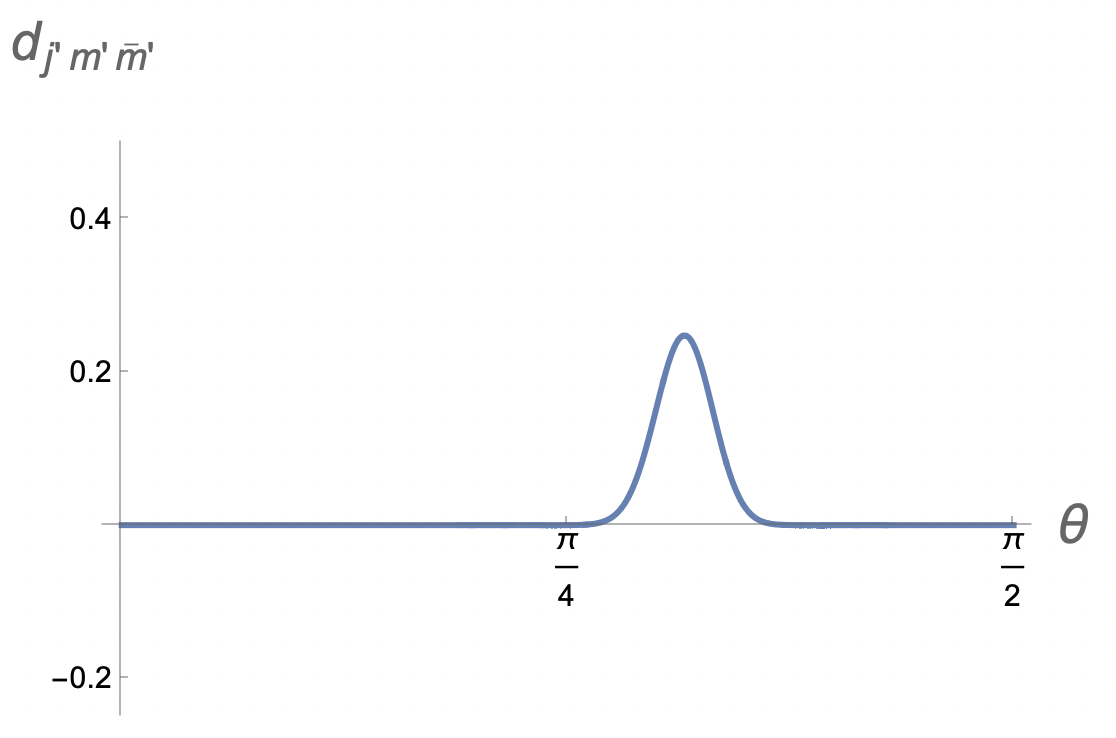}
    \caption{ }
    \label{fig:WignerD-extremal}
  \end{subfigure}
\hskip 2cm
  \begin{subfigure}[b]{0.38\textwidth}
  \vskip -2cm
    \includegraphics[width=\textwidth]{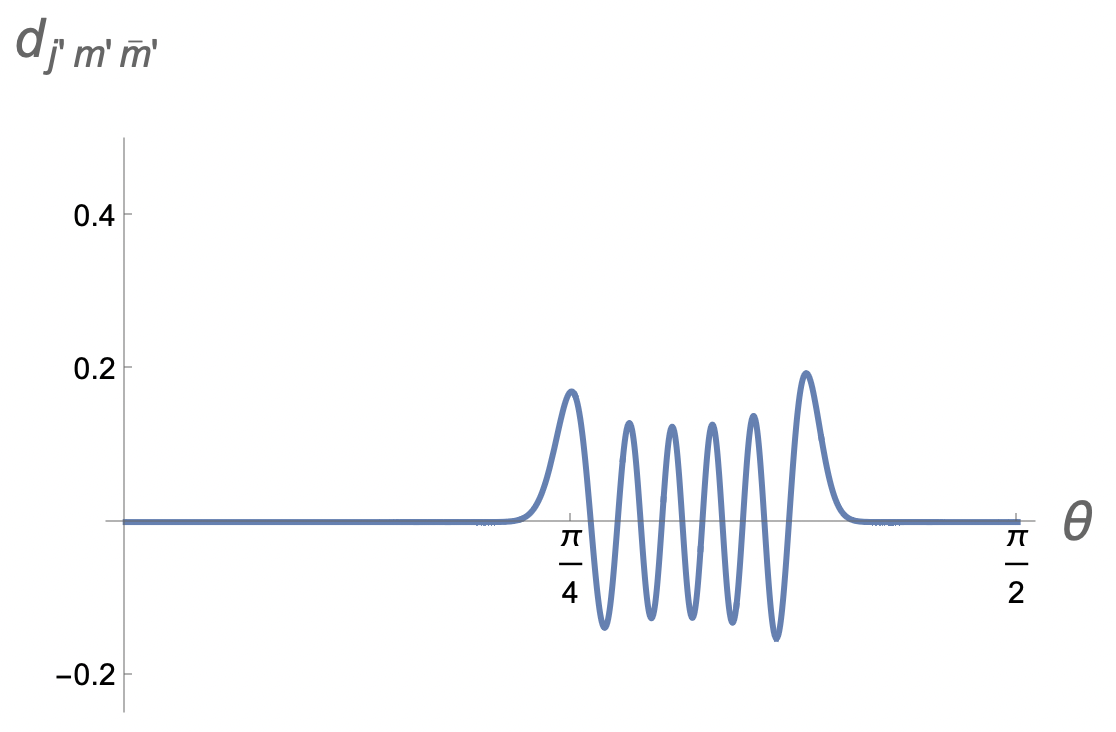}
    \caption{ }
    \label{fig:WignerD-generic}
  \end{subfigure}
\caption{ \it Wigner functions for $\sutwo$: 
(a) For $|\bar m'|=j'$ or $|m'|=j'$, the wavefunction is peaked at a particular polar angle $\theta$; the plot is for $|\bar m'|=j'$ and generic $m'$.
(b) For generic $|m'|,|\bar m'|<j'$, the wavefunction oscillates between minimum and maximum values (eq.~\eqref{thetapm} in the semi-classical limit), as the orbital axis is not aligned with the poles of the three-sphere.
}
\label{fig:Wgner fns}
\end{figure}

Figure~\ref{fig:Wgner fns} depicts two examples of these wavefunctions.  In Figure~\ref{fig:WignerD-extremal}, the BPS wavefunctions are stationary at a fixed value of $\theta$, while in Figure~\ref{fig:WignerD-generic}, the non-BPS wavefunctions describe particles oscillating back and forth in the effective potential of the $\sutwo$ Hamiltonian~\eqref{Heffs}.

The eigenfunctions of the scalar Laplacian on $AdS_3$ once again reflect the properties of the geodesics.  Formally, the $\bS^3$ metric analytically continues to that of $AdS_3$ (up to an overall sign) under a continuation of the $\bS^3$ polar coordinate $\theta$ to the $AdS_3$ radial coordinate; one can also check that the algebra of isometry generators continues from one to the other.  The discrete series representations of $\sltwo$ result from the continuation of the $\sutwo$ representations if we replace
\be
\label{sutosl}
j' \to -j
~~,~~~~
\msu \to \msl
~~,~~~~
\bmsu \to \bmsl
~~,~~~~
\mhat \to \nhat~,
\ee
in agreement with~\eqref{eq:mhat-nhat-defn}, and make the substitutions
\be
\label{coord subs}
(\coeff{\pi}2-\theta)\to -i\rho
~~,~~~~
\phi\to -\tau
~~,~~~~
\psi\to \sigma+\coeff\pi2 ~.
\ee
Then the lowest weight state wavefunction of the $\sutwo$ representation of spin $j'$ continues to the lowest weight state wavefunction of the positive discrete series $D^+_j$, 
\be
\label{hi wt map}
e^{-2ij'\phi}(\sin\theta)^{2j'} ~\longrightarrow~
e^{-2ij\tau}(\cosh\rho)^{-2j} ~~,
\ee
and the fact that the raising operators map from those of $\sutwo$ to those of $\sltwo$ guarantees that the rest of the wavefunctions~\eqref{su2wavefns} continue appropriately from $\sutwo$ to $\sltwo$.  The continuation of $j$ means that the representation has no highest weight (and therefore $j$ can be real rather than half-integer; the factorials $x!$ appearing in~\eqref{su2wavefns} should then be interpreted as $\Gamma(x+1)$). Upon quantization, the allowed range of $j$ is $1/2 < j < (\nfive+2)/2$, see \eg~\cite{Giveon:1998ns,Maldacena:2000hw,Argurio:2000tb,Martinec:2018nco} for discussions in the present context.

Similarly, if one starts near the $SU(2)$ highest weight state by setting $m=j-\nhat$, $\bar m=j-\nbarhat$ and performs the continuation~\eqref{sutosl}, one arrives near the highest weight state of the negative discrete series $D^-$.

Analogously to $\sutwo$, the extremal wavefunctions with either $\nhat=0$ or $\nbarhat=0$ have no nodes and are concentrated near $\rho_+=\rho_-$, see Eq.~\eqref{rhopm}, while the generic wavefunctions are standing waves which oscillate in the range $\rho_-\lesssim\rho\lesssim\rho_+$, and fall to zero over a scale $1/\sqrt{j}$ outside this range. 

%%%%%%%%%%%%%%%%%%%%%%%%%%%%%%%%%
\subsection{1/4-BPS wavefunctions in three-charge spectral flowed supertubes}
\label{sec:wavefuncinGLMT}

In the supersymmetric three-charge spectral flowed backgrounds, the wavefunctions must satisfy the conditions~\eqref{eq:geod-sf-2} that arose in the geodesic analysis. These relations constrain the choice of $AdS_3$ and $\bS^3$ wavefunctions.

Let us consider the BPS null geodesics discussed around \eqref{location}, and analyze the corresponding wavefunctions. Recall that to leading order in large quantum numbers, we have $j=j'$, $\bmsl=j$, $\bmsu=-j$, $\msl=j+\nhat$, $\msu=-j+\mhat$, and that the motion is stationary in $\rho, \theta, \phi, \psi$ at the location~\eqref{location}.
For these geodesics, from \eqref{eq:geod-sf-2} we have
\be
\label{eq:bps-geod-glmt}
\sfk \varepsilon \,=\, - 
\k n_y \;=\; 
\nhat + (2s+1) \mhat - 2 s j \;.
\ee
The corresponding wavefunctions take the form depicted in Fig.\;\ref{fig:WignerD-extremal}, 
with a peak at the values~\eqref{location} (and a width of order $j^{-1/2}$).
The resulting wavefunctions~\eqref{su2wavefns}, together with their $\sltwo$ counterparts, are none other than the mode functions $\Delta_{\ell \mhat\nhat}$ of the superstratum construction~\cite{Bena:2015bea,Bena:2016agb,Bena:2016ypk,Bena:2017xbt,Bena:2018bbd,Ceplak:2018pws,Heidmann:2019zws,Ganchev:2022exf,Ceplak:2022pep,Ceplak:2024dbj},
\begin{align}
\label{Deltafns}
\Delta_{\ell \mhat\nhat} \, e^{iv_{\ell \mhat\nhat}} = 
\bigg(\frac{r_\flat}{\sqrt{r_\flat^2+a^2}}\bigg)^\nhat\bigg(\frac{a}{\sqrt{r_\flat^2+a^2}}\bigg)^\ell\sin^{\ell-\mhat}\theta\,\cos^\mhat\theta \;
e^{i[ (\nhat+(2s+1)\mhat-2sj)v/\sfk+(\ell-\mhat)\phi-\mhat\psi ]} ~,
\end{align}
where $\ell=2j+1$
(and note the spacetime spectral flow from the group coordinates $\phi_{\rm NS},\psi_{\rm NS}$ to those of the GLMT background, Eq.~\eqref{NS2JMaRT}).
The dimensionless radial coordinate $r_\flat/a$ is related to our radial coordinate $\rho$ via (see Appendix~\ref{sec:asymflat})
\be
\label{rdef}
\frac{r_\flat^2}{a^2} = \sinh^2\rho ~.
\ee

We note that plugging \eqref{location} into \eqref{eq:bps-geod-glmt}, one obtains
\be
\label{eq:bps-geod-glmt-2}
\sfk \varepsilon \,=\, - 
\k n_y \;=\; 2j\Big( \sinh^2\rho + (2s+1)\cos^2\theta -s\Big) = \nfive\nu f_0 \;.
\ee
This is a generalization of the statement made earlier that on the locus $f_0=0$, massless BPS geodesics have $\varepsilon \,=\, - 
n_y \,=\,0\,.$ 

We recall that in the AdS$_3$ limit, the ergoregion is defined by $f_0<s(s+1)$, so the question of whether the energy of these geodesics is negative is not quite the same as the question of whether they are localised in the ergoregion.

Specifically, the negative energy massless BPS geodesics are located in the ergoregion, orbiting the $y$ circle at fixed $\rho, \theta, \phi, \psi$, in the region in which $f_0<0$.
By contrast, the massless BPS geodesics in the ergoregion at $f_0=0$ have zero energy, and those with $0<f_0<s(s+1)$ have positive energy.  This difference results from the fact that a massless particle must have spatial momentum and therefore energy to travel a stationary trajectory in the ergoregion.

More generally, let us consider including small non-zero values of $\alpha_r$, $\alpha'_r$, such that $\alpha_r < \alpha_\ell$ and $\alpha'_r<\alpha'_{\ell}$, and also allowing general values of the background spectral flow parameters $\sfm$ and $\sfn$.
Then combining Eqs.\;\eqref{thetapm}, \eqref{jmsu}, \eqref{rhopm}, \eqref{eq:m-alpha-l}, and \eqref{eq:geod-sf-2}, we obtain
\begin{align}
\begin{split}
\label{sugranull}
\sfk(\vareps-n_y) &\;\approx\; \nfive\nu\Big[ \cosh(\rho_+ \tight+ \rho_-) + (\sfm+\sfn)\cos(\theta_+ \tight+ \theta_-)\Big]~,
\\[.2cm]
\sfk(\vareps+n_y) &\;\approx \;\nfive\nu\Big[ \cosh(\rho_+ \tight- \rho_-) - (\sfm-\sfn)\cos(\theta_+ \tight- \theta_-)\Big]~. 
\end{split}
\end{align}
Adding the two equations in \eqref{sugranull}, we obtain
\be
\label{sugranull-1}
\sfk\vareps \;\approx\; \nfive\nu\Big[ \cosh\rho_+\cosh\rho_- -\sfm\sin\theta_+\sin\theta_- +\sfn\cos\theta_+\cos\theta_-  \Big] ~.
\ee

Returning to the supersymmetric backgrounds, for which $\sfm=s+1,\sfn=s$, one can expand the second equation in \eqref{sugranull} for small $\alpha_r$, $\alpha'_r$, to obtain
\begin{align}
\label{sugranull-2}
\sfk(\vareps+n_y) &\;\approx \;\frac{\nfive\nu}{2} \Big(\alpha_r^2+(\alpha'_r)^2\Big)~. 
\end{align}
Similarly for Eq.~\eqref{sugranull-1}, for $\sfm=s+1,\sfn=s$ and to leading order in small $\alpha_r$, $\alpha'_r$, one can interpret the products of trigonometric/hyperbolic functions as an approximate ``geometric mean'' of the two extremes of the massless geodesic, such that to leading order $\cosh\rho_+\,\cosh\rho_- \,\approx\, \ch^2\rho$, \etc. Then we recognize the expression in square brackets as again being $f_0$ rewritten using trigonometric identities,
\be
f_0 = \cosh^2\rho - (s+1)\sin^2\theta +s\,\cos^2\theta ~,
\ee
consistent with the discussion around Eq.~\eqref{eq:bps-geod-glmt-2} above.

%%%%%%%%%%%%%%%%%%%%%%%%%%%%%%%%%
\section{Supergravity spectrum, holography, and ergoregion modes}
\label{sec:sugra-spectrum}

In the fivebrane decoupling limit, the family of backgrounds \eqref{GLMTmetric2} admits an exact worldsheet description as a gauged Wess-Zumino-Witten model (WZW) for the group quotient  
\be
\label{cosets1}
\frac\cG\cH \,=\, \frac{\sltwo\times\sutwo\times \bR_\sft\times\bS^1_\sfy}{U(1)_L\times U(1)_R}~,
\ee
where $\mathcal{H}$ consists of a pair of null isometries of $\mathcal{G}$. 
Globally, we work with the universal cover of $\sltwo$ and we gauge a null cylinder, $\mathbb{R}\times S^1$.
There is also a $\bT^4$ factor that does not participate in the gauging. 

In this section, we first review the essential features of this construction, including the general structure of string vertex operators and the physical state constraints they obey. We then proceed to perform a more complete analysis of the spectrum of supergravity modes than has previously been performed in the literature, including the map to the dual spacetime CFT. We pay particular attention to the specific properties of supergravity ergoregion modes in subsection~\ref{sec:GLMTqrtrBPS}.

%%%%%%%%%%%%%%%%%%%%%%%%%%%%%%%%%%%%%%%%%%%%%%%%%%
\subsection{Worldsheet cosets and supergravity backgrounds}
\label{sec:backgds}

We use the Euler angle parametrization of the group $\sltwo$ and $\sutwo$ as given in \eqref{gslmat} and \eqref{gsumat1}. The line element for the (6+2)-dimensional part $\mathcal{G}$ of the (10+2)-dimensional target space before gauging is thus\footnote{We use the notation $\sft$, $\sfy$ for coordinates on the $\mathbb{R}\times S^1$ of the (10+2)-dimensional spacetime before gauging, to distinguish them from the $t,y$ coordinates of the (9+1)-dimensional spacetime after gauging.}
\be
\label{eq:upstairs-metric}
ds^2 = n_5 (-\cosh^2 \rhoo \:\! d\tau^2 + d\rho^2 + \sinh^2 \rhoo \:\! d\sigma^2) + n_5( d\theta^2 +  \cos^2\theta d\psi^2 + \sin^2\theta d\phi^2) - d\sft^2 + d\sfy^2 \,.
\ee
The null currents and their worldsheet superpartners are
\begin{align}
\begin{split}
\label{null-currents}
\cJ \,=\, J^3_\sl + l_2 J^3_\su + l_3 \,i\partial \sft + l_4\, i\partial \sfy
~~&,~~~~
\lambdab \,=\, \psi^3_\sl + l_2 \,\psi^3_\su + l_3 \,\psi^{\sft} + l_4\, \psi^\sfy ~,
\\[.2cm]
\bar\cJ \,=\, \bar J^3_\sl + r_2 \bar J^3_\su + r_3 \,i\bar\partial \sft + r_4\, i\bar\partial \sfy
~~&,~~~~
\bar\lambdab \,=\, \bar\psi^3_\sl + r_2 \,\bar\psi^3_\su + r_3 \,\bar\psi^{\sft} + r_4\, \bar\psi^{\sfy}  ~,
\end{split}
\end{align}
where the $J^3$, $\bar J^3$ currents were introduced in Eqs.\;\eqref{suqnums}, \eqref{slqnums}, and where
the parameters $l_i$, $r_i$ satisfy the null conditions
\be
n_5(-1 + l_2^2)- l_3^2 + l_4^2 \,=\, n_5(-1 + r_2^2)- r_3^2 + r_4^2 \,=\, 0  ~.
\ee

The general three-charge spectral flowed circular supertube solutions~\rcite{Giusto:2012yz}, in the NS5-brane decoupling limit  given in \eqref{GLMTmetric2}, correspond to the coset models with parameters~\rcite{Martinec:2017ztd,Martinec:2019wzw,Bufalini:2021ndn}\footnote{Compared to the conventions of~\rcite{Martinec:2022okx}, we have flipped the signs of $l_2,r_2$.}
\begin{align}
\label{lrglmt}
\begin{split}
l_2 = 2s+1
~,~~~~~
r_2=1
~,~~~~~
l_3 &= r_3 = -\Delta = -\bigg( \sfk R_y+\frac{n_5 s(s+1)}{\sfk R_y}\bigg) \;,
\\[.2cm]
l_4 = \sfk R_y - \frac{n_5 s(s+1)}{\sfk R_y}
~,~~~~~
r_4 &= -\sfk R_y - \frac{n_5 s(s+1)}{\sfk R_y} ~.
\end{split}
\end{align}
The quantity $\Sigma$ introduced in~\eqref{GLMTparams} is in general given by
\be\label{sigmanullgauging}
\Sigma \;=\; \sinh^2\rhoo + l_2 r_2 \cos^2\theta +\frac{1-l_2r_2}{2}+\frac{l_3r_3-l_4r_4}{2n_5} ~.
\ee

Gauge transformations act on the coordinates of the (10+2)-dimensional target space as
\begin{align}
\label{gaugetransfs}
\delta\tau & = l_1\alpha\tight+r_1\beta = (\alpha\tight+\beta) \;,\qquad\quad~~
\delta\phi = -l_2\alpha\tight-r_2\beta = -(\alpha\tight+\beta)(s+1) -(\alpha \tight-\beta)s \;,
\nn\\[1mm]
\delta\sigma &= l_1\alpha\tight+r_1\beta = (\alpha\tight-\beta)\;, \qquad\quad~~
\delta\psi =  l_2\alpha \tight-r_2\beta = (\alpha \tight+ \beta) s+(\alpha\tight-\beta) (s+1) \;,
\\[1.5mm]
\delta \sft &= l_3\alpha\tight+r_3\beta = -\,(\alpha\tight+\beta){\Delta}
\;,\qquad
\delta \sfy = { -l_4\alpha } 
\tight-r_4\beta =  {-\sfk R_y}\,(\alpha\tight-\beta)
+\frac{n_5 s(s+1)}{\sfk R_y}(\alpha\tight+\beta)\;.
\nn
\end{align}
We integrate out the gauge fields in the gauged sigma model action, which has the effect of adding a term $J\bar{J}/\Sigma$ to the sigma model action on $\mathcal{G}$. 
Upon fixing the gauge $\tau=\sigma=0$, the remaining coordinates become those of the physical target space background fields in Eq.~\eqref{GLMTmetric2}, with linear dilaton asymptotics.

Instead of fixing the gauge $\tau=\sigma=0$, one can define gauge-invariant coordinates of the physical spacetime after gauging. For ease of presentation, we focus on the leading form of these in the AdS$_3$ limit. 
We define $\tilde{\sft}=\sft/R_y\,,$ $\tilde{\sfy}=\sfy/R_y\,,$ analogously to \eqref{eq:ads3-limit}.
Then the gauge-invariant coordinates can be taken to be
\begin{align}
\begin{aligned}
\tilde{t}_\mathrm{gi} \;&=\; 
\tilde{\sft} + \sfk \tau ~, \qquad\qquad 
\tilde{y}_\mathrm{gi} \;=\; 
\tilde{\sfy} + \sfk \sigma ~, 
\end{aligned}
\end{align}
and for instance either 
\begin{align}
\label{gauge1}
\begin{aligned}
 {\phi}_{\mathrm{gi,1}}  \;&=\; \phi - \frac{s+1}{\sfk}\tilde{\sft} -  \frac{s}{\sfk} \tilde{\sfy} ~,
 \qquad\qquad 
{\psi}_{\mathrm{gi,1}}
\;=\; 
\psi +  \frac{s}{\sfk} \tilde{\sft}  +  \frac{s+1}{\sfk}\tilde{\sfy}  
 ~, \qquad\qquad 
\end{aligned}
\end{align}
or equally well,
\begin{align}
\label{gauge2}
\begin{aligned}
 {\phi}_{\mathrm{gi,2}}  \;&=\; \phi + {(s+1)}\tau +  s \sigma ~,
 \qquad\qquad 
{\psi}_{\mathrm{gi,2}}
&=
\psi -  s \tau -(s+1) \sigma  
~.
\end{aligned}
\end{align}
If we fix the gauge $\tau=\sigma=0$, we identify the respective upstairs coordinates $(\tilde{t}_{\mathrm{gi}},\tilde{y}_{\mathrm{gi}},\phi,\psi,\phi_{\mathrm{gi}},\psi_{\mathrm{gi}})$ with the downstairs coordinates $(\tilde{t},\tilde{y},\phi,\psi,\phi_{\NS},\psi_{\NS})$ in Eqs.~\eqref{eq:glmt-ads}--\eqref{eq:sf-GLMT-sugra}.

By contrast, if we instead gauge fix $\tilde{\sft}=\tilde{\sfy}=0$, then we identify the respective upstairs coordinates $(\tilde{t}_{\mathrm{gi}},\tilde{y}_{\mathrm{gi}},\phi=\phi_{\mathrm{gi}},,\psi=\psi_{\mathrm{gi}})$ with the downstairs coordinates $(\sfk \tau = \tilde{t},\sfk \sigma = {\tilde{y}},\phi,\psi)$ in Eqs.~\eqref{eq:glmt-ads}--\eqref{eq:sf-GLMT-sugra}. 
In this case, there is a residual discrete $\bZ_\sfk$ symmetry that acts as~\rcite{Martinec:2019wzw,Martinec:2023zha}
\begin{align}
\label{orbact}
\delta(\sigma,\psi,\phi) \;&=\; \frac{2\pi}{\sfk} \big( 1,s+1,-s \big) ~.
\end{align}
This residual symmetry corresponds directly to the coordinate identification that gives rise to the orbifold singularities in the supergravity solutions.

%%%%%%%%%%%%%%%%%%%%%%%%%%%%%%%%%%%%%%%%%%%%%%%%%%
\subsection{Holographic CFT states}

In the holographic CFT that arises in the AdS$_3$ limit, at the symmetric product orbifold point, the supersymmetric spectral flowed circular supertubes correspond to states in which all ``strands'' (\ie\ cycles in the symmetric orbifold twist sector) have winding $\sfk$ and are in the same state. For a review and more details on the notation used here, see~\cite{Avery:2010qw,Giusto:2012yz,Chakrabarty:2015foa}. The state on each $\sfk$-wound strand is a left spectral flow of the $\ket{++}_\sfk$ ground state, which we denote $\ket{++}_{\sfk,s}$.  It has the following quantum numbers under the holographic CFT $\mathsf{L}_0 , \mathsf{J}^3, \bar{\mathsf{L}}_0, \bar{\mathsf{J}}^3$ respectively~\rcite{Giusto:2012yz}$\,$:
\begin{align}
\ket{++}_{\sfk,s}: \qquad h &\,=\, \frac{\sfk}{4} + \frac{s(s+1)}{\sfk} \,, \qquad 
\Msu \,=\,  s+\frac{1}{2} \;, \qquad
\nonumber\\[.2cm]
\bar{h} &\,=\, \frac{\sfk}{4} \,, \qquad\qquad\qquad~~~ 
\bMsu \,=\,  \frac{1}{2} \;.
\label{eq:glmt-strands-1}
\end{align}
Compared to the ground state $\ket{++}_\sfk$, the spectral flowed state $\ket{++}_{\sfk,s}$ contains two Fermi seas of free fermions, filled to the level $s/\sfk$, with level spacing $1/\sfk$~\rcite{Giusto:2012yz}.

For later use in discussing the spectrum, we also define here the states 
\begin{align}
\ket{--}_{\sfk,s} \;=\; \ket{++}_{\sfk,s-1}: \qquad h &\,=\, \frac{\sfk}{4} + \frac{s(s-1)}{\sfk} \,, \qquad 
\Msu \,=\,  s-\frac{1}{2} \;, \qquad
\nonumber\\[.2cm]
\bar{h} &\,=\, \frac{\sfk}{4} \,, \qquad 
\bMsu \,=\,  \frac{1}{2} \;.
\label{eq:glmt-strands-2}
\end{align}
As a special case, $\ket{--}_{\sfk,s=1}=\ket{++}_\sfk$. For more general discussions of spectral flowed strands, see \eg~\rcite{Bena:2016agb,Shigemori:2022gxf}.

%%%%%%%%%%%%%%%%%%%%%%%%%%%%%%%%%%%%
\subsection{Physical string spectrum~-- general structure}
\label{sec:specreview}

In the null-gauged WZW model, the physical spectrum is the subsector of the underlying WZW model on $\cG$ satisfying the gauge constraints; in the application to worldsheet string theory, we also must impose the Virasoro constraints.  We first briefly review these constraints; for further details, see \rcite{Martinec:2018nco}.

Physical vertex operators are in the cohomology of the worldsheet BRST operator (for more details, see~\rcite{Martinec:2020gkv,Bufalini:2022wzu,Martinec:2022okx})
\be
\label{BRSToperator}
\cQ_{\rm\sst BRST} \,=\, \oint\!dz\, \big[ \big( c \;\! T + \gamma \;\! G +{\it ghosts} \big) + \big( \tilde c \;\! \cJ + \tilde\gamma \;\! \lambdab \big)\big] ~,  
\ee 
where $\cJ$, $\lambdab$ were defined in Eq.~\eqref{null-currents}.

To construct the vertex operators, one begins with the center-of-mass 
wavefunctions 
\be
\label{comfn}
\Phi^{(w)}_{j;m,\mbar} \,
\Psi^{(w',\bar w')}_{j';m',\mbar'} \,
e^{ -iE \sft + i P_y \sfy + i\bar P_{y} \bar \sfy } ~,
\ee
where $\Phi^{(w)}_{j;m,\mbar}$ is a primary of the bosonic $\sltwo$ WZW model in the spectral flow sector $w$;  $\Psi^{(w',\bar w')}_{j';m',\mbar'}$ is a bosonic primary of the $\sutwo$ WZW model in the (L,R) spectral flow sector $(w',\bar w')$; and $\sfy(z),\bar \sfy(\bar z)$ are the (anti-)chiral parts of the boson $\sfy$.%
\footnote{There is only a single spectral flow quantum number $w=\bar w$ for $\sltwo$ because we are working on the universal cover, so that there is no winding around the time direction~\rcite{Martinec:2018nco}.}
We have
\be
\label{pypybar}
P_y \,=\, \frac{n_y}{R_y} + w_y R_y ~, \qquad
\bar{P}_y \,=\, \frac{n_y}{R_y} - w_y R_y ~.
\ee
One then decorates this center-of-mass operator with $\sltwo$ and $\sutwo$ supercurrent raising operators as well as oscillators in the various free fields, \etc.

The total currents appearing for instance in~\eqref{null-currents} are given by
\be
\label{eq:J-full}
J_\sl^a = j_\sl^a - \frac i2 \T{(\epsilon_\sl)}{a}{bc}\psi_\sl^b \psi_\sl^c  
~~,~~~~
J_\su^a = j_\su^a - \frac i2 \T{(\epsilon_\su)}{a}{bc}\psi_\su^b \psi_\su^c  ~,
\ee
where $j^a_{\sl}$ and $j^a_{\su}$ are, respectively, bosonic $SL(2,\mathbb{R})$ level $n_5+2$ currents and bosonic $SU(2)$ level $n_5-2$ currents, $\epsilon_\sl^{123}=\epsilon_\su^{123}=1$, and indices are raised and lowered with the relevant Killing metric. We denote the total spins by $J,J'$ respectively, and by $M,M'$ the eigenvalues of the total $J^3_\sl,J^3_\su$ (and similarly for the right-movers).

We consider only vertex operators with vanishing $\bT^4$ momentum (see~\rcite{Martinec:2018nco} for its inclusion).
The zero-mode Virasoro constraints are~\rcite{Martinec:2018nco}%
\footnote{We adopt the conventions of~\rcite{Polchinski:1998rq} setting $\alpha'=1$, $X(z)X(w)\sim -\coeff{1}2 \log|z-w|^2$; note that $T(z)=-\partial X\partial X$.}
\begin{align}
\label{Vir-constr}
\begin{split}
L_0 - \frac12 
\,&=\,  -\frac{j(j-1)}{n_5} +
\frac{j'(j'+1)}{n_5}  - \Msl \wsl -
\frac{n_5}{4}\wsl^2 
+ \Msu \wsu + \frac{n_5}{4}\wsu^2
-\frac14 E^2 +\frac14 P_{y}^2 + h_L \,=\,0\;,  
\\[.3cm]
\bar L_0 - \frac12 \,&=\,  -\frac{\jsl(\jsl-1)}{n_5} +
\frac{\jsu(\jsu+1)}{n_5}  - \bMsl \bwsl -
\frac{n_5}{4}\bwsl^2 
+ \bMsu \bwsu+ \frac{n_5}{4}\bwsu^2
-\frac14 E^2 +\frac14 \bar{P}_{y}^2 + h_R \,=\,0 \;,  
\end{split}
\end{align}
where $h_{L,R}$ is the contribution of nonzero modes, plus the ground state energy $-1/2$.

It is convenient to rewrite the operators in a way that manifests the dependence on the total currents $J^3_\sl,J^3_\su$.  For $\sutwo$ one has
\be
\Psi^{(\wsu,\bwsu)}_{j';\msu,\bmsu} = \Lambda^{(\alpha_\su,\bar\alpha_\su)}_{j';\msu,\bmsu}\, \exp\biggl[i\frac{2}{\sqrt{\nfive}}\bigg(\Big(\Msu+\frac{\nfive}{2}\wsu\Big)Y' + \Big(\bMsu+\frac{\nfive}{2}\bwsu\Big)\bar Y' \bigg) \biggr] ~,
\ee
where $Y',\bar Y'$ bosonize the total $J^3$ currents,
\be
J^3_\su = i\sqrt{\nfive}\, \partial Y'
~~,~~~~
\bar J^3_\su = i\sqrt{\nfive}\, \bar\partial \bar Y' \;,
\ee
and where $\Lambda$ is a {\it super-parafermion} operator that lives in the $\sutwo/\uone$ coset model, see e.g.~\rcite{Martinec:2018nco,Martinec:2020gkv,Balthazar:2021xeh}.  The parameters $\alpha_\su,\bar\alpha_\su$ are the $\cR$-charge spectral flow quantum numbers of the super-parafermion under its $\cN=2$ worldsheet superconformal symmetry, and roughly codes the $\psi^\pm_\su$ fermion charge (and so note that one has the relations $\Msu\tight=\msu\tight+\alpha_\su,\bMsu\tight=\bmsu\tight+\bar\alpha_\su$).
The operator $\Lambda$ has conformal dimension
\be
\label{hpf}
h_\Lambda = \frac{j'(j'+1)}{\nfive} - \frac{\Msu\:\!{}^2}{\nfive} + \frac{\alpha_\su^2}{2} ~,
\ee
(which is positive, by unitarity);
thus the $\Psi$ conformal dimension can be written as
\be
h_\Psi = h_\Lambda + \frac1{\nfive}\Big( \Msu+\frac{\nfive}2\wsu\Big)^2 \,.
\ee
The zero-mode null constraints then read
\begin{align}
&(2\Msl + n_5 \wsl) + l_2 (2\Msu+n_5
    \wsu)  + l_3 E  +  l_4 P_{y} \;=\;0 ~ ,  \nn \\
&(2 \bMsl + n_5 \bwsl) + r_2 (2\bMsu+ n_5 \bwsu) + r_3 E  +  r_4 \bar{P}_{y} \;=\;0 ~ . \label{null-constr}
\end{align}

The supergraviton vertex operators are constructed by decorating the center-of-mass vertex operator~\eqref{comfn} with one excitation each on the left and right, in order to satisfy the GSO projection.
For zero winding $\wsl=\wsu=\bwsu=w_y=0$, the Virasoro constraints~\eqref{Vir-constr} require $\jsl=\jsu+1$.

The circular supertube background, which is a special case of the above backgrounds in which $s=0$, was studied in~\rcite{Martinec:2020gkv,Martinec:2022okx}. 
The null constraints and the BPS condition impose a lowest weight condition $\Msl=J=J'=-\Msu$ on the left, and similarly on the right, on the total $\sltwo$ and $\sutwo$ spins.  These vertex operators,  denoted 
\be \label{eq:half-bps-vert}
\cV^{\alpha \dot{\alpha}}, \qquad \cS^{AB} ~,
\ee
sew together $2\jsu\tight+1$ background strands~\eqref{eq:glmt-strands-1} (with $s=0$) into a single strand of length $(2\jsu\tight+1)\sfk$, while changing the 1/2-BPS ground state polarization in a manner determined by that of the vertex operator,
\be
\label{eq:hol-cft-ST-1}
\big(\ket{++}_{\sfk}\big)^{2j'+1} ~~\longrightarrow ~~
\ket{I}_{(2j'+1)\sfk} ~.
\ee
These operators have a 1/4-BPS generalization, in which one relaxes the lowest weight condition on the total spins $\Msl,\Msu$.  These operators sew together background strands in the same way, while in addition populating the resulting strand with excitations by 
$\sfJ_{-\frac{1}{\sfk}}^+$ and
$\sfL_{-\frac{1}{\sfk}}\tight- \sfJ^3_{-\frac{1}{\sfk}}$.  These modes, and various generalizations where we allow arbitrary rather than BPS polarization on the left, are the linearized perturbations associated to superstratum deformations of the round supertube background~\rcite{Martinec:2022okx}.

%%%%%%%%%%%%%%%%%%%%%%%%%%%%%%%%%%%%%%%%%%
\subsection{1/4-BPS supergravity spectrum}
\label{sec:GLMTqrtrBPS}

We now discuss the generalization of these results to the GLMT backgrounds. We begin with an analysis of the supergravity spectrum, in which we have $w=w'=\bar{w}'=0$. In the NS sector in the $(-1)$ picture, we have $h_L=h_R=0$, and in the R sector in the $(-1/2)$ picture, we have a ground state spin field on both sides. 

The axial Virasoro constraint, from~\eqref{Vir-constr}, imposes  
\begin{equation}
\label{eq:axialvir-glmt-sugra}
n_y w_y =0 \, .
\end{equation}
We solve this by setting $w_y=0$. Note that the states with $n_y=0$ and nonzero $w_y$ are not supergravity modes in this duality frame, but become supergravity modes after T-dualizing along $y$; for more discussion, see~\rcite{Martinec:2022okx}.
We work in a large $R_y$ expansion, and write
\be
\label{eq:E-wy-eps}
E = \frac{\vareps}{R_y} ~.
\ee
Then $\vareps$ has the interpretation of energy measured with respect to $\partial/\partial \tilde{t}$ in the rotating $AdS_3$ backgrounds \eqref{eq:glmt-ads}.

Focusing first on the NS-NS sector, at leading order in large $R_y$ the vector combination of the Virasoro constraints~\eqref{Vir-constr} imposes
\be
\label{jjprel}
j \,=\, j'+1 \;.
\ee
This relation receives corrections away from the AdS$_3$ limit~\rcite{Martinec:2018nco}, but we will work in the AdS$_3$ limit from now on.

We consider positive discrete series states in $SL(2,\mathbb{R})$, which correspond to modes that have positive frequency with respect to the AdS ``cap'' geometry in factorized coordinates~\rcite{Martinec:2018nco}.
In this sector, at leading order in large $R_y$, the zero-mode left and right null constraints become
\begin{subequations}
\label{lr-glmt-null}
\begin{align}
\sfk(\vareps-n_y) &\,=\, 2\Msl  + 2 (2s+1) \Msu \, , \label{l-null-glmt-ads} \\
\sfk(\vareps+n_y) &\,=\, 2 \bMsl  + 2 \bMsu  \, . \label{r-null-glmt-ads}
\end{align}
\end{subequations}
Taking the sum and difference of these equations, we obtain the zero-mode vector and axial null constraints, 
\begin{subequations}
\label{va-glmt-null}
\begin{align}
\sfk\vareps &\,=\, \Msl+\bMsl + (2s+1) \Msu + \bMsu \, , \label{v-null-glmt-ads}  \\
-\sfk n_y &\,=\, \Msl- \bMsl + (2s+1) \Msu - \bMsu  \, . \label{a-null-glmt-ads}
\end{align}
\end{subequations}
We note that in the semiclassical approximation, one does not distinguish $\Msl$ from $\msl$, and similarly for $\bMsl$, $\Msu$, $\bMsu$. Then we see that the zero-mode vector and axial null constraints agree with the geodesic analysis of the previous section, Eq.~\eqref{eq:geod-sf-2}.

To analyze the 1/4-BPS spectrum, we solve the BPS constraint on the right by fixing 
\begin{equation}
\label{GLMTmbmbp}
\bar{M} = \bar{J} = j+\bar{\epsilon} \, ,\qquad \bar{M}' = -\bar{J}' = -(j'+\bar{\epsilon}')
\end{equation}
where $\bar\epsilon$, $\bar\epsilon'$ denote the contributions to the total spins from the worldsheet fermions in the vertex operator, and are taken to be either $\bar\epsilon=-1, \bar\epsilon'=0$ (denoted $\bar{\cV}^+=\bar{\cW}^-$ in~\rcite{Martinec:2022okx}) or by taking  $\bar\epsilon=0, \bar\epsilon'=1$ (denoted $\bar{\cV}^-=\bar{\chi}^+$ in~\rcite{Martinec:2022okx}). 
In both cases, we have $\bar\epsilon'-\bar\epsilon=1$ which implies that $\bMsl=-\bMsu$. Then the R null constraint, Eq.~\eqref{r-null-glmt-ads},
imposes 
\be \label{eq:E-minus-ny}
\vareps \,=\, - {n_y} \,.
\ee
Note that we are dealing with positive frequency modes of left-moving excitations, so the typical regime of parameters of most excitations is $n_y<0$ and $E>0$\,. By contrast, we will also exhibit a finite set of modes that have $n_y>0$ and $E<0$\,, which are examples of ergoregion modes, c.f.~the discussions in Section \ref{sec:ergoregion} and below Eq.~\eqref{eq:geod-sf-2}. Note that positive frequency modes with $E<0$ are not present in the two-charge supertube backgrounds.
These modes reduce the absolute value of both the energy and $y$-momentum from the background, while preserving supersymmetry. 

We finally solve the remaining null constraint, which we can take to be the axial null constraint. From \eqref{lrglmt} and \eqref{null-constr}, and using $\bar{M}=-\bar{M}'$, this reads:
\begin{equation}
\label{Lnull GLMT}
M + (2s+1)M' + \sfk n_y 
\,=\,0 \; . 
\end{equation}
On the left we define
\begin{equation} \label{eq:mhat-nhat}
M = J + \nhat = (j+\epsilon) +\nhat \,, \qquad M' = -J' + \mhat = - (j'+\epsilon')+\mhat \, ,
\end{equation}
with $\nhat = 0, 1, 2 \dots$ and $\mhat = 0, 1, 2,\dots, 2J'$; again $\epsilon$, $\epsilon'$ denote the contributions to the total spins from the worldsheet fermions in the vertex operator. 
In the holomorphic NS sector, these can take values within $\{-1,0,1\}$, with more freedom than in the anti-holomorphic sector (discussed below \eqref{GLMTmbmbp}), as we shall describe in more detail momentarily. Together with \eqref{GLMTmbmbp}, the axial null constraint then becomes
\begin{equation}
\label{eq:axialnull-glmt}
\nhat + (2s+1) \mhat - 2 s j' +\epsilon - (2s+1)\epsilon' +1 = -\sfk n_y \,.
\end{equation}
Since $n_y \in \mathbb{Z}$, the left-hand side of this equation must be a multiple of $\sfk$ in order for the vertex operator to be physical.

Note that in the semiclassical approximation, this equation reduces to Eq.~\eqref{eq:bps-geod-glmt} which arose in the analysis of the corresponding BPS geodesics. The following is therefore the quantum version of the analysis of the semiclassical modes in Section \ref{sec:wavefuncinGLMT}.

To explain the structure step by step, let us first examine \eqref{eq:axialnull-glmt} with $\nhat=\mhat=0$.  Generically, such vertex operators will not satisfy $n_y \in \mathbb{Z}$, however this condition will be satisfied in due course by restoring $\nhat$ and $\mhat$. We obtain
\begin{equation}
 - 2 s j' +\epsilon - (2s+1)\epsilon' +1 \,=\, -\sfk n_y  \;.
\end{equation}
As an example, let us choose the holomorphic $\epsilon$ variables similarly to those in the antiholomorphic sector, either by taking $\cV^+=\cW^-$ which has $\epsilon=-1, \epsilon'=0$ or by taking $\cV^-=\chi^+$ which has $\epsilon=0, \epsilon'=1$.
Combined with the antiholomorphic sector, this leads to the four operators $\cV^{\alpha\dot\alpha}$. 
Then we are left with 
\begin{equation}
\label{eq:ergo-modes-glmt-1}
 - 2 s j'  - 2s\epsilon'  \,=\, -\sfk n_y  \;,
\end{equation}
where again $\epsilon'=0$ for $\cV^+$ and $\epsilon'=1$ for $\cV^-$. 
For simplicity, let us temporarily assume that $s/\sfk \in \mathbb{Z}$, to ensure that we solve the constraint that $n_y \in \mathbb{Z}$.

Recall that we work with $s\ge 0$ without loss of generality. 
For $s=0$, there is no ergoregion; the lightest excitations are 1/2-BPS, have $\varepsilon=n_y=0$ and $w_y \geq 0$, and mediate transitions to other two-charge BPS states~\rcite{Martinec:2020gkv,Martinec:2022okx}.
For $s>0$ and for $j'>1$, the vertex operators in \eqref{eq:ergo-modes-glmt-1} have $n_y>0$, which is in the opposite sense to the background, and thus reduce the absolute value of the momentum charge $n_y$ compared to its value of the background. Moreover, from \eqref{eq:E-minus-ny}, these operators have negative asymptotic energy, despite being positive frequency modes with respect to the local cap energy (being $D^+$ modes).

The two terms in \eqref{eq:ergo-modes-glmt-1} can be understood from the holographic CFT, as follows. 
First, let us consider $\cV^{++}$, so that $\epsilon'=0$. We then have $n_y = 2 j's/\sfk $. The relevant process in the holographic CFT is (c.f.~Eqs.~\eqref{eq:glmt-strands-1}--\eqref{eq:glmt-strands-2})
\be
\label{eq:hol-cft-glmt-1}
\big(\ket{++}_{\sfk,s}\big)^{2j'+1} ~~\longrightarrow ~~
\ket{++}_{(2j'+1)\sfk \;\! , \;\! (2j'+1)s} \;.
\ee
In more detail, this vertex operator joins together $2j'+1$ strands of length $\sfk$. The original strands of length $\sfk$ had two Fermi seas filled to the level $s/\sfk$, as discussed around Eq.~\eqref{eq:glmt-strands-1}. Following the discussion in~\cite[Sec.~6]{Giusto:2012yz}, 
the change in $m'$, $\bar{m}'$ in this process is fully accounted for by the polarization of the ground states, and the total number of fermionic excitations is preserved. The lowest available state on the new strand of length $(2j'+1)\sfk$ is two filled Fermi seas, each containing $(2j'+1)s$ fermions filling the Fermi sea to the level $s/\sfk$, with energy spacing $[(2j'+1)\sfk]^{-1}$. The reduction in chiral energy $h$ (and therefore the reduction in the absolute value of $n_y = \bar{h}-h$) is precisely $2  j's/\sfk $, as found in~\cite[Eq.~(6.22)]{Giusto:2012yz}.\footnote{Note that compared to the notation of \rcite{Giusto:2012yz}, $(2j'+1)_{\mathrm{here}}=l_{\mathrm{there}}$.}

Next, we consider $\cV^{--}$, so that $\epsilon'=1$. The new element with respect to $\cV^{++}$ is the term $-2s\epsilon' = -2s$ in Eq.~\eqref{eq:ergo-modes-glmt-1}. The relevant transition in the holographic CFT is 
\be
\label{eq:hol-cft-glmt-2}
\big(\ket{++}_{\sfk,s}\big)^{2j'+1} ~~\longrightarrow ~~
\ket{--}_{(2j'+1)\sfk \;\! , \;\! (2j'+1)s} ~=~ \ket{++}_{(2j'+1)\sfk \;\! , \;\! (2j'+1)(s-1)} \;.
\ee
In this example, the reduction in the absolute value of $n_y$ is precisely $2  s(j'+1)/\sfk $, again in agreement with \eqref{eq:ergo-modes-glmt-1}.

The discussion above generalizes straightforwardly to the polarizations $\ket{+-}$ and $\ket{-+}$, via the vertex operators $\cV^{+-}=\cV^{+}\bar{\cV}^{-}$ and $\cV^{-+}=\cV^{+}\bar{\cV}^{-}$, following the discussion around Eq.\;\eqref{eq:half-bps-vert}.

We next generalize the discussion to include the quantum numbers $\mhat$, $\nhat$, and relax the temporary assumption $s/\sfk \in \mathbb{Z}$, to return to working with general non-negative integer $s$. The worldsheet vertex operators are those that fill out the $\sltwo$ and $\sutwo$ multiplets, \cf~\eqref{eq:mhat-nhat}. We identify the corresponding operators in the holographic CFT as being those that mediate the transitions to the family of superstratum-type states built on spectrally flowed strands, with final states ($I=\alpha\dot\alpha=\pm \pm$) 
\begin{align}
\label{origstrata}
\big| \mhat,\nhat;I\big\rangle_{(2j'+1)\sfk\;\! , \;\! (2j'+1)s} \;=\; \big(\sfJ_{-\frac{(2s+1)}{\sfk}}^+ \big)^\mhat \big( \sfL_{-\frac{1}{\sfk}}- \sfJ^3_{-\frac{1}{\sfk}} \big)^{\nhat}  \big| I\big\rangle_{(2j'+1) \sfk\;\! , \;\! (2j'+1)s} \;.
\end{align}
We note in particular the fractional mode numbering of the raising operators on the right-hand side, which is consistent with the axial null constraint~\eq{eq:axialnull-glmt}.
The fractionation is limited in the sense that the modes are multiples of $1/\sfk$, whereas the modes on the strand are fractionated in multiples of $1/((2j'+1)\sfk)$. This is a generalization of the (limited) fractionation seen in the supergravity constructions and holographic analysis of~\rcite{Bena:2016agb,Shigemori:2022gxf}.  The holographic dictionary for superstrata is on a firm footing, having passed precision holographic tests~\rcite{Giusto:2015dfa,Giusto:2019qig,Rawash:2021pik}.

The ``supercharged'' superstratum modes~\rcite{Ceplak:2018pws,Heidmann:2019zws} now follow from generalizing the discussion in~\rcite{Martinec:2022okx}. Specifically, we study the vertex operators that differ from $\cV^{\pm}$ by changing the signs of $\epsilon$ and $\epsilon'$. 
We first discuss changing the sign of $\epsilon'$, namely $\epsilon=0,\epsilon'=-1$, keeping the right-movers unchanged. In the case $\bar{\epsilon}=0,\bar{\epsilon}'=1$, the axial null constraint becomes
\begin{equation}
\label{eq:axialnull-glmt-2}
\nhat + (2s+1) \mhat - 2 s j' + (2s+2) \,=\, -\sfk n_y  \; .
\end{equation}
This family of vertex operators corresponds to holographic CFT operators that mediate transitions to the superdescendant states
\begin{align}
\Big(\sfG^{+1}_{-\frac{s+1}{\sfk}} 
\sfG^{+2}_{-\frac{s+1}{\sfk}}
+\frac{1}{\sfk} \sfJ^+_{-\frac{2s+1}{\sfk}} \big(\sfL_{-\frac{1}{\sfk}}-\sfJ^3_{-\frac{1}{\sfk}}\big)\Big)\, \big|\mhat,\nhat;++\big\rangle_{(2j'+1)\sfk \;\! , \;\! (2j'+1)s} ~.
\end{align}

Alternatively, we can change the sign of $\epsilon$, namely $\epsilon=+1,\epsilon'=0$, keeping the R movers unchanged. In the case $\bar{\epsilon}=-1,\bar{\epsilon}'=0$, the axial null constraint becomes
\begin{equation}
\label{eq:axialnull-glmt-3}
\nhat + (2s+1) \mhat - 2 s j' +2 \,=\, -\sfk n_y 
 \, .
\end{equation}
This corresponds to transitions to 
\begin{align}
\Big(\sfG^{+1}_{-\frac{s+1}{\sfk}} 
\sfG^{+2}_{-\frac{s+1}{\sfk}}
+\frac{1}{\sfk} \sfJ^+_{-\frac{2s+1}{\sfk}} \big(\sfL_{-\frac{1}{\sfk}}-\sfJ^3_{-\frac{1}{\sfk}}\big)\Big)\, \big|\mhat,\nhat;--\big\rangle_{(2j'+1)\sfk \;\! , \;\! (2j'+1)s} \;.
\end{align}
Again, the ground state polarizations $\ket{+-}$ and $\ket{-+}$ follow entirely analogously.

We note that the moding of $\sfJ^+_{-\frac{(2s+1)}{\sfk}}$ agrees with the observations on the supergravity spectrum in~\cite[Eq.~(6.22)]{Giusto:2012yz}. The modings of $\sfJ^+_{-\frac{(2s+1)}{\sfk}}$ and $\sfG^{+A}_{-\frac{(s+1)}{\sfk}}$ also agree with the transformation of the generators under spectral flow, 
\begin{equation}
\label{eq:hol-cft-glmt-n}
\sfL_{\frac{n}{\sfk}} \rightarrow \sfL_\frac{n}{\sfk} \, ,\quad~~ \sfJ^{\pm}_{\frac{n}{\sfk}} \rightarrow \sfJ^{\pm}_{\frac{n\mp 2s}{\sfk}} \, ,\quad~~ J^3_{\frac{n}{\sfk}}\rightarrow \sfJ^3_{\frac{n}{\sfk}} \, , \quad~~ \sfG^{\pm,A}_{\frac{n}{\sfk}} \rightarrow \sfG^{\pm,A}_{\frac{n\mp s}{\sfk}} \;,
\end{equation}
as discussed in the perturbative analysis of general superstrata on spectral flowed supertube backgrounds in~\rcite{Shigemori:2022gxf}.

So far we have explicitly identified 8 NS-NS vertex operators for each allowed choice of $j',\mhat, \nhat$. There are another 8 NS-NS massless 1/4-BPS vertex operators that lie in six-dimensional vector multiplets. These arise from combining a BPS vertex $\bar{\cV}$ for the right-movers with a $\mathbb{T}^4$ polarization $\cZ^{A\dot{A}}$ of the left-movers. 
In addition, there are 16 R-R massless 1/4-BPS vertex operators for each allowed choice of $j',\mhat, \nhat$. The analysis of these remaining vertex operators follows directly from combining the analysis above with that in~\cite[Sec.\;5.2]{Martinec:2022okx}.

%%%%%%%%%%%%%%%%%%%%%%%%%%%%%%%%%%%%%%%
%%%%%%%%%%%%%%%%%%%%%%%%%%%%%%%%%%%%%%%
%%%%%%%%%%%%%%%%%%%%%%%%%%%%%%%%%%%%%%%

\section{Stringy spectrum and twisted sector ground states}
\label{sec:stringspec}

In this section we consider more general string states. We begin with a general analysis of the spectrum of strings with nonzero winding along $\bS^1_y$. The gauge constraints relate winding along $\bS^1$ and in $\sltwo$; furthermore, one can shift the winding from one to the other via a large gauge transformation.  The quantities that can be compared with conserved charges in the holographically dual spacetime CFT should be gauge-invariant, and so we construct the gauge-invariant worldsheet currents which measure these charges. 
This leads to an analysis of winding sector strings which extend the supergravity vertex operators to untwisted sectors of non-zero winding.
We then analyze twisted sector strings that are pinned to the orbifold singularities of the $\sfk>1$ backgrounds.
One of the orbifold singularities lies in the ergoregion, and we exhibit ergoregion modes in this sector.
We also exhibit ergoregion modes among a class of massive string states.

%%%%%%%%%%%%%%%%%%%%%%%%%%%%%%%%%%%%
\subsection{Stringy spectrum: Constraints at large \texorpdfstring{$R_y$}{}}
\label{sec:constraints}

We now discuss a subset of states with non-zero $w,w',\bar{w}',w_y$.
To do so, when $w_y \neq 0$, and at large $R_y$, we write 
\be
\label{Eexpn}
E = w_y R_y + \frac{\vareps}{R_y}
\;,
\ee
where $w_y$ and $\vareps$ are independent of $R_y$.
Combining this with the expressions for $P_y$, $\bar{P}_y$ in \eqref{pypybar}, we find that at leading order in large $R_y$, the zero-mode L and R null constraints are respectively 
\begin{align}
\begin{split}
\label{genlargeRnull}
\sfk(\vareps-n_y) &= 2\Big(\Msl_\tot +\frac{\nfive}{2}\wsl\Big) + (2s+1)2\Big(\Msu_\tot+\frac{\nfive}{2}\wsu\Big) -\frac{2s(s+1)}{\sfk}n_5w_y \;,
\\[.2cm]
\sfk(\vareps+n_y) &= 2\Big(\bMsl_\tot+\frac{\nfive}{2}\wsl\Big) + 2\Big(\bMsu_\tot+\frac{\nfive}{2}\bwsu\Big)   \;,
\end{split}
\end{align}
where we have defined
\be
\label{mtotdef}
\Msl_\tot=\Msl+\Msl_\osc
~~,~~~~
\Msu_\tot=\Msu+\Msu_\osc \;,
\ee
and similarly for $\bMsl_\tot,\bMsu_\tot$, where  $\Msl_\osc,\Msu_\osc$ are the corresponding contributions of the oscillator excitations charged under $J^3_\sl,J^3_\su$, and similarly for $\bMsl_\osc,\bMsu_\osc$. 
The large-$R_y$ limit of the Virasoro constraints~\eqref{Vir-constr} is similarly
\begin{align}
\begin{split}
\label{largeRvir}
0 &= \frac{-j(j-1)+j'(j'+1)}{n_5} - \Msl_\tot\wsl + \Msu_\tot\wsu-\frac{n_5}4\Big(\wsl^2\tight-(\wsu)^2\Big) - \frac{w_y}{2}\big(\vareps \tight- n_y \big)
+ h^{~}_L ~,
\\[.3cm]
0 &= \frac{-j(j-1)+j'(j'+1)}{n_5} - \bMsl_\tot\wsl+\bMsu_\tot\bwsu-\frac{n_5}4\Big(\wsl^2\tight-(\bwsu)^2\Big) - \frac{w_y}{2}\big(\vareps \tight+ n_y \big) 
+ h^{~}_R ~,
\end{split}
\end{align}
where $h^{~}_L,h^{~}_R$ contain the non-zeromode excitation levels and the super-parafermion $\cR$-charge spectral flows,
\be
h^{~}_L \,=\, N_L + \frac{\alpha_\sl^2+\alpha_\su^2}{2} - \half
~~,~~~~
h^{~}_R \,=\, N_R + \frac{\bar\alpha_\sl^2+\bar\alpha_\su^2}{2} - \half  ~.
\ee
Although we will focus exclusively on the above constraint zero modes, we note that one must also satisfy all the positive frequency modes of the constraints as well.  We will assume that the set of solutions to these additional constraints is non-empty.

%%%%%%%%%%%%%%%%%%%%%%%%%%%%%%%%%%%%
\subsection{Gauge-invariant charges}
\label{sec:charges}

When turning on $\sfy$-circle winding $w_y \neq 0$, one must be careful to define conserved charges such as energy, momentum, and angular momentum in a gauge invariant way.
The various $\uone$ currents of the null-gauged model are 
\be
J^3_\sl ~,~~ 
J^3_\su ~,~~ 
i\:\!\partial \sft\equiv J_t ~,~~ 
i\:\!\partial \sfy\equiv J_y ~~;~~~~
\bar J^3_\sl ~,~~
\bar J^3_\su ~,~~
i\:\!\bar\partial \sft \equiv \bar J_t ~,~~
i\:\!\bar\partial \sfy \equiv \bar J_y ~,
\ee
which measure the respective zero-mode quantum numbers
\be
\begin{split}
\Msl_\tot+\frac{n_5}2 w ~&,~~ 
\Msu_\tot+\frac{n_5}2 \wsu ~,~~ 
\frac E2 ~,~~ 
\half\Big(\frac{n_y}{R_y} + w_y R_y\Big) ~;~~ 
\\[.2cm]
\bMsl_\tot+\frac{n_5}2 w ~&,~~
\bMsu_\tot+\frac{n_5}2 \bwsu ~,~~
\frac E2 ~,~~ 
\half\Big(\frac{n_y}{R_y} - w_y R_y\Big) ~~.
\end{split}
\ee
To be physical, currents must commute with the null currents $\cJ,\bar\cJ$ in~\eqref{null-currents}, and must measure combinations of zero-mode quantum numbers that are invariant under the following shifts induced by the combination of spectral flow transformations which amount to a large gauge transformation, as discussed in~\rcite{Martinec:2018nco},
\begin{align}
\begin{aligned}
\label{gauge specflow}
\delta w = q ~~,~~~~
\delta w' = -l_2 q ~~&,~~~~
\delta \bar w' = -r_2 q ~~,
\\[.2cm]
\delta E = l_3 q ~~,~~~~
\delta P_y = -l_4 q ~~&,~~~~
\delta \bar{P}_y = -r_4 q ~~,
\quad~~~ 
q\in\bZ ~,
\\[.2cm]
\Longrightarrow \quad
\delta n_y \;=\; -\nfive \frac{s(s+1)}{\sfk} q \;&,
\quad 
\delta w_y \;=\; -\sfk q \;.
\end{aligned}
\end{align}

For instance, recalling that $l_3=r_3$, we have the following examples of physical currents:
\begin{align}
\label{Mcurrents}
\cM\;&=\;J^3_\sl - \frac{n_5}{l_3} J_t
~~,~~~~~~~~
\bar\cM\;=\; \bar J^3_\sl - \frac{n_5}{l_3} \bar J_t ~, 
\\[.2cm]
\label{Mpcurrents}
\cM'\;&=\; J^3_\su - \frac{\nfive l_2}{l_4} J_y
~~,~~~~
\bar\cM'\;=\; \bar J^3_\su - \frac{\nfive r_2}{r_4} \bar J_y ~,
\\[.2cm]
\cL\;&=\; 
J_t + \frac{l_3}{l_4} J_y
~~,~~~~~~~~~
\bar\cL\;=\;
\bar J_t + \frac{l_3}{r_4} \bar J_y
\label{Lcurrents}
~.
\end{align}
These are not all independent~-- for instance the linear combination
\be
\cM + l_2 \cM' + \frac{(l_3^2+\nfive)}{l_3} \cL\,=\, \cJ
\ee
is trivial upon imposing the constraints.

The combination of zero mode quantum numbers measured by the current $\cM-\bar\cM$ is the cap AdS$_3$ angular momentum,
\be
\cM-\bar\cM \,=\, \Msl_\tot-\bMsl_\tot ~,
\ee 
where we use the same notation to refer to both the current and its eigenvalue.
At leading order in large $R_y$, the combination of zero-mode quantum numbers measured by $\cM_\tot+\bar\cM_\tot$ is
\be
\cM+\bar\cM \,=\, \Msl_\tot+\bMsl_\tot+\nfive\left(w+\frac{w_y}{\sfk}\right) \;,
\ee
which is a gauge-invariant non-zero-winding generalization of the cap energy of supergravity modes, $\Msl_\tot+\bMsl_\tot$.

At leading order in large $R_y$, the combination of zero mode quantum numbers measured by  $\cL+\bar{\cL}$ is 
\be
\label{epstil value}
\cE \;=\; \varepsilon - 
n_5 \frac{s(s+1)}{\sfk}
\frac{w_y}{\sfk}
\ee
which is a gauge-invariant non-zero-winding generalization of the asymptotic energy $\varepsilon$ of supergravity modes, in which the energy of wound strings is compared to a reference energy per unit winding of the background, times the winding.
Similarly, the leading term in the large $R_y$ expansion of the zero-mode quantum numbers measured by $\bar\cL-\cL$ 
is 
\be
\label{nytil value}
\cN_{\sst\!\cY} \;=\; n_y + 
n_5 \frac{s(s+1)}{\sfk}\frac{w_y}{\sfk}
~,
\ee
and the currents $\cM',\bar\cM'$ measure the $\bS^3$ angular momenta%
\footnote{While it might appear that this expression violates angular momentum quantization in twisted sectors $w_y\notin \sfk\bZ$, this is an artifact of the definition, in which the angular momentum of individual strings is compared to the angular momentum per unit winding of the background.  In the background, each strand of the holographic CFT accounts for $\sfk$ units of winding and $(s+\half)$ units of angular momentum (using the same labeling as the symmetric product orbifold), \cf~Eq.~\eqref{eq:glmt-strands-1}, and so the angular momentum per unit winding is $(s+\half)/\sfk$.  In the worldsheet theory, each fundamental string winding translates into $\nfive$ units of winding in CFT terms, accounting for the fractional term in~\eqref{cM value}.  Note that if we have several such strings whose windings $w_y$ add up to a multiple of $\sfk$, as would for instance happen if a unit of background flux were converted into several fractionally wound strings, the total angular momentum would in fact be properly quantized.}
\be
\label{cM value}
\cM' = \Msu_\tot + \frac{\nfive}2
\Big(
\wsu-(2s+1)\frac{w_y}\sfk
\Big)
~~,~~~~
\bar\cM' = \bMsu_\tot + \frac{\nfive}2
\Big(
\bwsu- \frac{w_y}\sfk
\Big)
\ee
respectively.
One can verify that the quantities $\cM$, $\bar\cM$, $\cM'$, $\bar\cM'$, $\cL$ and $\bar\cL$  are indeed invariant under the spectral flow large gauge transformations~\eqref{gauge specflow}.

Recall that when one uses the gauge symmetry to fix the $U(1)_\sfy$ coordinate of the gauged WZW model, 
there is a residual discrete gauge symmetry that arises~\rcite{Martinec:2023zha}, given in~\eqref{orbact}, that corresponds to the supergravity identification~\eqref{eq:sugra-orbact-GLMT}.  When considering the effect of large gauge transformations, one can shift away $w_y$ in multiples of $\sfk$, see Eq.~\eqref{gauge specflow}.%
\footnote{Alternatively, one can use the gauge spectral flow symmetry to choose a gauge $w=0$, and then $w_y$ accounts for all of the string winding charge.}
We can thus restrict to the range 
\be
\label{twistsec}
w_y\,\in\,\{0,1,...,\sfk-1\}   ~,
\ee
and use $w$ to account for string winding in multiples of $\sfk$, which are integer units of F1 winding, while $w_y$ can be thought of as labeling orbifold twisted sectors.  We can then regard $w_y$ as labeling twisted sectors under the $\bZ_\sfk$ orbifold~\eqref{orbact}, implemented by simultaneous ``fractional spectral flow'' in $\sltwo$ and $\sutwo$.

%%%%%%%%%%%%%%%%%%%%%%%%%%%%%%%%%%%%
\subsection{Solving the constraints}
\label{sec:consolve}

When the windings $w,w_y$ are nonzero, one can solve the constraints~\eqref{genlargeRnull}, \eqref{largeRvir} for $\vareps\pm n_y,\,\nhat=\Msl\!-\!J,\,\nbarhat=\bMsl\!-\!\bar J$, as defined in \eqref{eq:mhat-nhat}. 
When doing so, it is useful to express the solutions to the constraints in terms of gauge-invariant combinations.

Although there are solutions to the constraints in nonzero winding sectors involving continuous series representations of $\sltwo$, these are unbound winding strings in plane wave states and thus not localized in the cap.  Thus, we continue to focus on solutions for which the $\sltwo$ center of mass wavefunction is in the discrete series $D^{+}$.

We remind the reader that $n_y,\nhat,\nbarhat$ must be integers, and that $\nhat,\nbarhat$ must be non-negative. For the left-moving quantities, one finds 
\begin{align}
\label{lefties-e}
\cE-\cN_{\sst\!\cY} 
&~=~ 
\frac{2}{\sfk w+w_y}\Bigg[\frac{-j(j-1)+j'(j'+1)}{\nfive} +h_L 
\nn\\[.1cm]
&\hskip 2.5cm
+ \Msu_\tot\Big(\wsu-(2s+1)\frac{w_y}{\sfk}\Big) +  (2s+1)\Msu_\tot\Big(\wsl+\frac{w_y}{\sfk}\Big) 
\\[.1cm]
&\hskip 2.8cm
+ \frac\nfive4\bigg(\Big(\wsl+\frac{w_y}{\sfk}\Big)^2 + \Big(\wsu-(2s+1)\frac{w_y}{\sfk}\Big)^2 
\nn\\[.1cm]
&\hskip 4cm
+2(2s+1)\Big(\wsl+\frac{w_y}{\sfk}\Big)\Big(\wsu-(2s+1)\frac{w_y}{\sfk}\Big) \bigg)
\Bigg] \quad{}
\nn
\end{align}
and
\begin{align}
\label{lefties-n}
\nhat &~=~ \frac{\sfk}{\sfk w+w_y}\bigg[\frac{-j(j-1)+j'(j'+1)}{\nfive} +h_L \nn\\[.1cm]
&\hskip 3.0cm
-(J+\Msl_\osc)\Big(\wsl+\frac{w_y}{\sfk}\Big) + \Msu_\tot\Big(\wsu-(2s+1)\frac{w_y}{\sfk}\Big)
\\[.1cm]
&\hskip 4cm
+\frac\nfive4\bigg(- \Big(\wsl+\frac{w_y}{\sfk}\Big)^2 + \Big(\wsu-(2s+1)\frac{w_y}{\sfk}\Big)^2 \bigg)
\bigg]  ~.
\nn
\end{align}
Similarly, for the corresponding right-moving quantities, one finds 
\begin{align}
\label{righties-e}
\cE+\cN_{\sst\!\cY} 
&~=~ \frac{2}{\sfk w+w_y}\Bigg[\frac{-j(j-1)+j'(j'+1)}{\nfive} +h_R 
+ \bMsu_\tot\Big(\bwsu-\frac{w_y}{\sfk}\Big) + \bMsu_\tot\Big(\wsl+\frac{w_y}{\sfk}\Big) 
\\[.1cm]
&\hskip 2.8cm
+ \frac{\nfive}{4}\bigg(\Big(\wsl+\frac{w_y}{\sfk}\Big)^2 + \Big(\bwsu-\frac{w_y}{\sfk}\Big)^2 
+2\Big(\wsl+\frac{w_y}{\sfk}\Big)\Big(\bwsu-\frac{w_y}{\sfk}\Big) \bigg)
\Bigg] \quad{}
\nn
\end{align}
and
\begin{align}
\label{righties-n}
\nbarhat &~=~ \frac{\sfk}{\sfk w+w_y}\bigg[\frac{-j(j-1)+j'(j'+1)}{\nfive} +h_R 
-(\bar J+\bMsl_\osc)\Big(\wsl+\frac{w_y}{\sfk}\Big) + \bMsu_\tot\Big(\bwsu-\frac{w_y}{\sfk}\Big)
\nn\\[.1cm]
&\hskip 3cm
+ \frac\nfive4\bigg(-\Big(\wsl+\frac{w_y}{\sfk}\Big)^2 + \Big(\bwsu-\frac{w_y}{\sfk}\Big)^2  \bigg)
\bigg]  ~.
%\nn
\end{align}
We have grouped the terms on the right hand sides of~\eqref{lefties-e}--\eqref{righties-n} into combinations that are invariant under the gauge spectral flow shifts, as described in Eq.~\eqref{gauge specflow}.  These expressions show explicitly how $w_y$ enters as a fractional string winding related to orbifold twist sectors.
The expressions~\eqref{lefties-e}, \eqref{righties-e} generalize the supergravity mode analysis of Section~\ref{sec:GLMTqrtrBPS} to general string modes.  Note that the quantization and non-negativity of $\nhat,\nbarhat$ constrain the choices on the RHS of Eqs.~\eqref{lefties-n}, \eqref{righties-n}.

We can now explore the stringy ergoregion modes by extending the logic of Section~\ref{sec:GLMTqrtrBPS}, where we now allow $\sutwo$ spectral flow as well as more general oscillator excitations, for any of the twist sectors labeled by $\sfw=w+w_y/\sfk$ (including the untwisted sector $w_y=0$).  With negative $\sutwo$ spectral flows $w',\bar w'$, we can arrange for an arbitrarily negative asymptotic energy of the vertex operator in sectors of large $y$-circle winding~-- the energy cost of $\sutwo$ spectral flow is less on longer strands, and the angular momentum overall is less, and so after the transition the energy is less than that of the initial state.  We will now give a class of examples that illustrate this phenomenon.

%%%%%%%%%%%%%%%%%%%%%%%%%%%%%%%%%%%%
\subsection{Winding ergo-strings}
\label{sec:ergo-wound}

We now investigate winding sector strings that are localized in the ergoregion.  In the zero winding sector, string vertex operators implement transitions built upon the basic process~\eqref{eq:hol-cft-glmt-1}, \eqref{eq:hol-cft-glmt-2} in the spacetime CFT.  The natural winding sector generalization of this transition is (specializing to $\sfw=w\tight+w_y/\sfk\in\bZ$, and maintaining the polarization state, for simplicity)
\be
\label{hol-cft-wound}
\big(\ket{++}_{\sfk,s}\big)^{2j'+1+n_5\sfw} ~~\longrightarrow ~~
\ket{++}_{(2j'+1+n_5\sfw)\sfk \;\! , \;\! (2j'+1+n_5\sfw)s} ~,
\ee
with possible additional excitations needed on the long strand on the RHS, in order to ensure integer momentum and angular momentum, as we shall discuss presently.  The above transition generalizes an analysis of BPS winding strings in the unflowed backgrounds with $s=0$ in~\rcite{Martinec:2020gkv,Martinec:2022okx}.
The change in conserved charges resulting from such a transition is given by
\begin{align}
\begin{aligned}
\label{delta qnums}
\Delta J^3_\su &\,=\, 
\Delta \bar J^{3}_\su \,=\, 
-\Big(j'+\frac{n_5}{2}\sfw\Big)~,
\\[.2cm]
\Delta L_0 &\,=\, 
-\Big(2j'+n_5\sfw\Big)\frac{s}{\sfk} ~, \qquad
\Delta \bar L_0 \,=\, 0  ~.
\end{aligned}
\end{align}
Note that the final state $L_0$ is in general not an integer; this means that one must excite some fractionally moded oscillators on the left in order to have properly quantized $y$-momentum.

This difference is accounted for by the charges of the vertex operator.  
As mentioned at the start of this section, charges in the spacetime CFT should be matched to the corresponding gauge-invariant charges $\cM',\bar\cM',\cE\pm\cN_{\sst\!\cY}$ of Eqs.~\eqref{epstil value}--\eqref{cM value} (in particular, this is why we have used the gauge-invariant quantity $\sfw$ in the above expressions).

Consider first the leading $n_5$-dependent terms.  To match the angular momenta~\eqref{cM value} and~\eqref{delta qnums} at leading order, we should choose
\be
\label{wsu match}
\wsu = 2s\frac{w_y}{\sfk} - w
~~,~~~~~~~
\bwsu = - w ~,
\ee
and then one finds for the terms proportional to $\nfive$ 
\be
\label{L0 match}
\frac{\cE-\cN_{\sst\!\cY}}{2}\bigg|_{\propto n_5} = -n_5 \frac{s}{\sfk}\Big( w+\frac{w_y}{\sfk}\Big)
~~,~~~~~~~
\frac{\cE+\cN_{\sst\!\cY}}{2}\bigg|_{\propto n_5} = 0 ~, ~~~
\ee
in agreement with~\eqref{delta qnums}.  Note that with the choices~\eqref{wsu match}, the terms in $\nhat,\nbarhat$ that are leading order in large $n_5$ cancel, and one has 
\begin{align}
\label{nnbar general}
\nhat &= \frac{\sfk}{\sfk w+w_y}\bigg[\frac{-\jsl(\jsl-1)+\jsu(\jsu+1)}{\sfk}+h_L\bigg] -\big(J+M_{\osc}+\Msu_{\tot}\big) ~,
\nn\\[.2cm]
\nbarhat &= \frac{\sfk}{\sfk w+w_y}\bigg[\frac{-\jsl(\jsl-1)+\jsu(\jsu+1)}{\sfk}+h_R\bigg] -\big(J+M_{\osc}+\Msu_{\tot}\big) ~.
\end{align}
For the $\sltwo$ ground state, there are no charged oscillator contributions, and we have $\jsl=\jsu+1$, $h_L=h_R=0$, and $\Msu=\bMsu=-J$.  Excitations must arrange that the two quantities $\nhat,\nbarhat$ are non-negative integers.

Again setting~\eqref{wsu match}, the energy and $y$-momentum~\eqref{lefties-e}, \eqref{righties-e} of a general string vertex operator are
(here and in the following, we ignore the subtle distinctions among the various polarization states of the string, \cf~\eqref{GLMTmbmbp}, \eqref{eq:mhat-nhat}, which make minor modifications to the expressions due to the distinction between $J$ and $j$, \etc.) 
\begin{align}
\label{eps ny general}
\cE &= 2\bigg(\frac{-\jsl(\jsl-1)+\jsu(\jsu+1)}{\nfive(\sfk w+w_y)}\bigg)
+\frac{h_L+h_R}{\sfk w+w_y} +\Big(\frac{s}{\sfk}\Big)\Big[2\Msu_\tot - \nfive\Big(w+\frac{w_y}{\sfk}\Big)\Big]
\nn\\[.2cm]
\cN_{\sst\!\cY} &= -\,\frac{h_L-h_R}{\sfk w+w_y} - \Big(\frac{s}{\sfk}\Big)\Big[2\Msu_\tot - \nfive\Big(w+\frac{w_y}{\sfk}\Big)\Big]  ~;
\end{align}
again, the quantum numbers must be chosen so that $n_y\in\bZ$. 

Turning off all the excitations, and setting $\Msu=\bMsu=-j$, one finds an exact match to~\eqref{delta qnums}, beyond the leading order already observed in~\eqref{L0 match}.  In addition, just as we saw in the spacetime CFT, one must add some number of fractionally moded left-moving excitations in order to have integral $y$-momentum.

We see that there is a reservoir of negative energy coming from the last two terms in $\cE$, that increases with increasing worldsheet spectral flow. This reservoir can be tapped to compensate the energy of excitations on the string (which are increasingly less costly at higher winding due to momentum fractionation), while still lowering the energy of the system.

%%%%%%%%%%%%%%%%%%%%%%%%%%%%%%%%%%%%
%%%%%%%%%%%%%%%%%%%%%%%%%%%%%%%%%%%%

\subsection{Twisted sector ground states}
\label{sec:twistgd}

A special case of stringy modes are the ground states in the twisted sectors of the orbifold identification~\eqref{orbact}.  These should be unexcited strings ($h_L=h_R=0$) that are pinned to the orbifold singularities of the background.  Because the background is rotating, a fractionally wound string might not be able to exactly co-rotate, because this would require a fractional angular momentum value that is not allowed.  In such cases, the best one can do is approximate with the closest integer value of (angular) momentum, and such a string generically must have some oscillator excitation.

The structure of the orbifold singularities in the background depends on the diophantine relations among the integers $(s,s+1,\sfk)$~\rcite{Jejjala:2005yu,Giusto:2012yz}.  The product $s(s+1)$ must be an integer multiple of $\sfk$ in order to avoid pathologies,%
\footnote{In the dual CFT, the momentum per strand must be in integer; in the worldsheet theory, the axial gauge orbits must close; in supergravity, there should be no horizons, closed timelike curves or singularities beyond orbifolds~\rcite{Bufalini:2021ndn}.}
so we write
\be
s+1 = \ell_1 m_1
~~,~~~~
s = \ell_2 m_2
~~,~~~~
\sfk = \ell_1\ell_2 ~,
\ee
for positive integers $\ell_1,\ell_2,m_1,m_2$.
The geometry has a $\bZ_{\ell_1}$ singularity at $\rho=0,\,\theta=0$ where $y$ and $\phi$ degenerate, and a $\bZ_{\ell_2}$ orbifold singularity at $\rho=0,\,\theta=\pi/2$ where $y$ and $\psi$ degenerate, see Eqs.~\eqref{eq:glmt-ads}--\eqref{eq:sugra-orbact-GLMT} and the analysis in~\rcite{Giusto:2012yz}.

The right-moving null and Virasoro constraints are satisfied for BPS states, for which
\be
\jsl=\jsu+1
~~,~~~~
\bMsl=J=J'= -\bMsu
~~,~~~~
\wsl = -\bwsu 
~~,~~~~
\bMsl_\osc=\bMsu_\osc=0
~.
\ee
We then look for left-moving ground states with cap energy/momentum $\nhat=0$, and no oscillator excitations.  There are two sets of such states; in the spectral flow gauge~\eqref{twistsec}, the first set has
\be
\label{set1}
w_y = p \ell_1
~~,~~~~
\wsl = \wsu = p m_2
~~,~~~~
\Msu= -J' 
~~,~~~~
\vareps = -n_y = -\frac{2J'm_2}{\ell_1}
~~,~~~~
p=1,...,\ell_2\tight-1 ~,
\ee
where we require $J'\in\ell_1\bZ$ to ensure $n_y\in\bZ$. The second set consists of
\be
\label{set2}
w_y = -p \ell_2
~~,~~~~
\wsl = -\wsu =  p m_1
~~,~~~~
\Msu = J' 
~~,~~~~
\vareps = -n_y = \frac{2J'm_1}{\ell_2}
~~,~~~~
p=1,...,\ell_1\tight-1 ~,
\ee
where this time we impose $J'\in\ell_2\bZ$.

Note that for large $J'$, 
the first set, having $\Msu=\bMsu=-J'$, is localized at $\theta=\pi/2$, in the deepest part of the ergoregion at the supertube locus;
similarly, for large $J'$ the second set, having $\Msu=-\bMsu=J'$, is localized at $\theta=0$.%
\footnote{The string center-of-mass wavefunctions were exhibited in Section~\ref{sec:wavefns}.  Note that worldsheet spectral flow, spinning up the string and extending it along the Euler angles $\tau,\sigma,\phi,\psi$, does not change its location in $\rho,\theta$.}  

Note that in twisted sectors of the orbifold, the gauge invariant charges of the vertex operator are fractional, because they are being compared to a fraction of a background string winding, so matching the vertex operator to a specific transition in the spacetime CFT is intrinsically a bit imprecise. Here we interpret the effect of the vertex operator as creating the long string with the quantum numbers it specifies, together with an additional residue of short twisted strings, left over from the integer number of initial state background strands one needs to sew together to form the long fractionally wound string.  These additional short twisted strings are in general excited.

The quantum numbers~\eqref{set1} are compatible with a vertex operator that implements the transition
(for example considering a R-R vertex operator polarization)
\be
\label{hol-cft-twisted2}
\big(\ket{++}_{\sfk,s}\big)^{2j'+n_5pm_2+\ceil{n_5p/\ell_2}} ~\longrightarrow ~
\ket{\Upsilon_1}
\ket{00}_{(2j'+\nfive pm_2)\sfk+n_5p\ell_1 \;\! , \;\! 2j's+n_5pm_2(s+1)} ~,
\ee
where $\ket{\Upsilon_1}$ consists of a set of short twisted strands that account for the small mismatch in the strand budget between the in-state and the long strand in the out-state, and has excitations that account for the small mismatch in $J^3_\su,\bar J^3_\su$ and $L_0$.

Similarly, the quantum numbers of~\eqref{set2} are compatible with a vertex operator that implements the transition 
\be
\label{hol-cft-twisted1}
\big(\ket{++}_{\sfk\, ,\,s}\big)^{2j'+n_5pm_1-\floor{n_5p/\ell_1}} ~~\longrightarrow ~~
\ket{\Upsilon_2}
\ket{00}_{(2j'+\nfive pm_1)\sfk-n_5p\ell_2 \, , \, 2j'(s+1)+n_5 pm_1s} ~;
\ee
again the state $\ket{\Upsilon_2}$ accounts for the small mismatch in quantum numbers.

For the twisted sector strings not in these two series, the term in $\nhat$ proportional to $\nfive$ is nonzero, and so generically such strings are located away from the cap of the geometry by an amount of order the AdS scale.

%%%%%%%%%%%%%%%%%%%%%%%%%%%%%%%%%%%%
\subsection{Non-winding massive string spectrum}
\label{sec:stringy}

We can also find stringy ergoregion modes in the sector where all worldsheet spectral flows and windings are equal to zero, $\wsl=\wsu=\bwsu=w_y=0$.  
The vector Virasoro constraint, from~\eqref{largeRvir}, then
yields
\begin{align}
\label{josc}
j &= \half\Big( 1+\sqrt{(2j'+1)^2+2n_5(h_L+h_R)} \,\Big) ~.
\end{align}
The axial Virasoro constraint sets $h_L=h_R$.
We see that there are discrete series solutions up to levels of order $\sqrt{n_5}$ without violating the unitarity bound for discrete series representations,
\be
\label{unidisc}
\half < j < \frac{n_5+1}2 ~.
\ee
The null constraints without winding are~\eqref{va-glmt-null}; we can again investigate the existence of ergoregion modes whose $\sltwo$ center-of-mass excitation is in the positive discrete series $D^+$, so that the cap energy $\Msl+\bMsl>0$, while the asymptotic energy is negative, $\vareps<0$.  
In parallel with the supergravity analysis of Section~\ref{sec:sugra-spectrum}, we can arrange such modes by starting with sufficiently large and negative $\sutwo$ angular momenta $\Msu,\bMsu$.

We see that there is scope for oscillator excitations while remaining in the unitary range of $j$, Eq.~\eqref{unidisc}, for oscillator levels up to of order $N_{L,R}\sim n_5/4$.
The inclusion of oscillator modes increases $j$ without increasing $j'$, and so leaves less scope for the possibility that the asymptotic and cap energies have opposite signs when solving the constraints.  Thus, in addition to the supergravity sector, there will be a spectrum of ergoregion modes among the unwound stringy excitations (a Hagedorn spectrum at sufficiently high level).

The cost of oscillator excitations is clear from their scale relative to the $AdS$ scale.  The $AdS$ curvature radius is $\sqrt\nfive$ times the string scale, and so we see that in the energy budget, a string oscillator costs of order $\sqrt\nfive$ times that of a momentum excitation on either $AdS_3$ or $\bS^3$.

%%%%%%%%%%%%%%%%%%%%%%%%%%%%%%%%%%%%%%%%%%%%
%%%%%%%%%%%%%%%%%%%%%%%%%%%%%%%%%%%%%%%%%%%%
\section{Summary and discussion}
\label{sec:disc}

In this work, we have explored the spectrum of fundamental strings in three-charge supersymmetric rotating $(AdS_3\times\bS^3)/\bZ_\sfk$ backgrounds which correspond to 1/4-BPS states of the holographically dual CFT.
These supergravity solutions have the feature of a supersymmetric ergoregion, without an event horizon.
The presence of the ergoregion means that there are supersymmetric excitations that lower the asymptotic energy while preserving the BPS relation between energy and charges.

We have used the exact worldsheet description of these backgrounds in terms of gauged Wess-Zumino-Witten models to study their spectrum of perturbative string excitations, at the non-perturbative level in $\alpha'$.
We have exhibited both massless and massive string states that are localized in the ergoregion, and that can lower the energy as measured from infinity, or exactly preserve it. 

The ungauged theory has two time coordinates, $\tau\in\sltwo$ and $\tilde\sft\in\bR_\sft$, that are related by the gauge constraints.  The utility of such a description is that, to a first approximation, $\tilde\sft$ is the asymptotic time that corresponds to the time in the holographic CFT, while $\tau$ is the time in the natural co-rotating frame in the cap of the geometry. 
While the two are tied together by the gauge constraints, they serve as useful heuristics in the corresponding regions of the geometry where most of the physical time coordinate flows along one rather than the other variable.

In previous work, we analyzed the 1/4-BPS supergravity spectrum on two-charge (1/2-BPS) backgrounds~\rcite{Martinec:2022okx}, finding a match to the set of linearized superstratum excitations (for a review, see~\rcite{Shigemori:2020yuo}).  However, this analysis was restricted to vertex operators in the sector of zero worldsheet spectral flow.  We have extended these results in two directions.  In Section~\ref{sec:GLMTqrtrBPS}, we analyzed the 1/4-BPS supergravity spectrum in three-charge supersymmetric spectral flowed (GLMT) backgrounds, which map onto superstratum deformations of these backgrounds~\rcite{Shigemori:2022gxf}. 
In Section~\ref{sec:ergo-wound}, these results were extended to sectors of non-zero worldsheet spectral flow.  A feature of these non-zero winding sectors is that, for $s\ne0$ and $\sfk>1$, momentum quantization generically forces such vertex operators to excite left-moving oscillations above the ground state, and so lie outside of supergravity.

For all of these 1/4-BPS vertex operators, we identified the corresponding transitions they implement in the dual spacetime CFT. 
The construction in Section~\ref{sec:charges} of gauge-invariant charges on the worldsheet provided a crucial ingredient in building this map in winding sectors, since these are the worldsheet quantities that match onto the difference of conserved charges between the initial and final states. 

Another feature of these backgrounds is the generic presence of orbifold singularities at two loci in their core, one of which is inside the ergoregion. We identified twisted sector ground state ergo-strings that are localized at this orbifold singularity, and candidates for the transitions they engineer in the dual CFT, in Section~\ref{sec:twistgd}.

Now that we have an improved understanding of the spectrum of perturbations of these backgrounds, one can re-examine the question of whether and how the resulting dynamics leads to scrambling of such a perturbation, such that the system evolves to a more typical state. 
The dynamics depends on whether the system is coupled to the ambient flat space, such that the bound state can radiate away angular momentum via low-energy supergravitons, or whether we consider the system in AdS with Dirichlet boundary conditions, such that the superselection sector is fixed.
The dynamics when the system is coupled to flat space has been previously considered~\rcite{Eperon:2016cdd} (see also~\cite{Marolf:2016nwu}), and clarified in~\cite{Bena:2018mpb,Martinec:2020gkv}, so here we focus on working at a fixed superselection sector after the initial perturbation.

The vertex operators that mediate transitions that lower the energy of the system do so by lowering the $\bS^3$ angular momentum $J^3_\su$, since the energy of the background comes from left-moving $\sutwo$ spectral flow of the spacetime background (or equivalently, the spacetime CFT).
By contrast, when working in a fixed superselection sector and thus demanding that the momentum and angular momentum remain unchanged, vertex operators have non-negative energy. 
In order not to change the net charges of the system, the vertex operator must have vanishing left/right $\sutwo$ charges, as well as momentum along $\bS^1_y$.  This requires $\Msu_\tot=\wsu=\bMsu_\tot=\bwsu=n_y=0$.
From the solution~\eqref{lefties-e}--\eqref{righties-n} of the constraints, the lower bounds $\nhat, \nbarhat\ge 0$ then force a positive $h_{L,R}$ in order to compensate a negative contribution from the winding $w$ in the cap energy.  Looking then at the asymptotic energy, the contributions of $w$ and $h_{L,R}$ are both positive, and so the effect of the vertex operator is to raise the energy of the state. This is perhaps not a surprise, as spectral flow is the most energetically efficient mechanism for imparting a given charge to a system on a single strand.

However, in a fixed superselection sector for momentum and angular momentum, there is the possibility of multi-particle (or more generally multi-string) processes that can maintain or lower the energy. 
For $\sfk>1$, the holographic CFT contains more entropic sectors of states with the same charges as the backgrounds we have considered.  
For instance, there is a phase with one long excited strand and several short ground state strands of winding less than $\sfk$ (e.g.~all of winding 1).
The short strands account for the bulk of the $\bar J^3$ angular momentum, and the long strand accounts for the $y$-momentum and the bulk of the $J^3$ angular momentum. The long strand gives rise to entropy from the number of ways of partitioning a given total $y$-momentum among highly fractionated oscillators~\cite{Bena:2011zw} (see also the related recent work~\cite{Larsen:2025jqo}).

So we can consider a process in which we assemble a subset of $q$ background strands of length $\sfk>1$ into a similar configuration of one long excited strand and several short ground state strands, while keeping angular momenta and $y$-momentum fixed, making sure to respect the condition of integer momentum per strand.
Since the long strand excitations are more highly fractionated, the transition frees up energy which can be distributed in various ways among the more fractionated oscillators, leading to a density of final states that grows with the length of the long strand.

What is the likelihood that the system can make such a transition? One can first join together a number of initial strands into a virtual strand that carries the combined angular momentum and is thus non-BPS.
The amplitude will be suppressed by the energy cost of the virtual state, for the time it takes to shed its angular momentum onto shorter strands, but enhanced by the larger (potentially Hagedorn-like) phase space of final states.  The computation of the amplitude for such transitions provides an interesting avenue of future investigation. 
For related work, see~\cite{Mathur:2008kg,Kraus:2015zda}.

This represents a specific string-theoretic process, in both the bulk and dual CFT descriptions, by which the system can evolve to more typical states when perturbed. It differs from that considered in~\rcite{Eperon:2016cdd} in that it takes place in the decoupled theory, rather than coupling the throat to an asymptotically flat spacetime region. Moreover, the analysis of~\rcite{Eperon:2016cdd} is done for the $\sfk=1$ backgrounds, whereas it is important in the above discussion that $\sfk>1$.

The fact that there are processes in the decoupled theory which make use of ergo-strings by shedding angular momentum onto a collection of supergravity modes in the throat indicates that, when one backs slightly away from the decoupling limit, and the geometry opens out onto asymptotically flat spacetime, there can be superradiant processes in which a string scattering off the throat carries away energy, $y$-momentum and angular momentum, leaving behind a BPS throat of reduced energy.

The present work also opens up opportunities for studying other features of the string spectrum, including for instance giant gravitons.
In related upcoming work~\cite{BDMSTW}, a family of vertex operators is studied, that describe supergravity waves and wound fundamental strings carrying momentum and angular momentum in supersymmetric two- and three-charge backgrounds. 
These strings precisely match solutions to the bubble equations of multi-center supergravity solutions.

Finally, in a companion paper to the present work, we shall carry out a parallel analysis of strings in the JMaRT family of non-supersymmetric three-charge backgrounds~\cite{MMT-6}.  Again there is an ergoregion, and supergravity ergoregion modes, but without supersymmetry. This makes a substantial difference~-- when the fivebrane throat is matched onto asymptotically flat spacetime, these non-supersymmetric backgrounds have a linear instability whereby they decay exponentially rapidly via the emission of quanta that carry away momentum and angular momentum.  This phenomenon has been well studied in the emission channel of massless scalar quanta~\rcite{Cardoso:2005gj,Chowdhury:2007jx,Chowdhury:2008bd,Chowdhury:2008uj,Avery:2009tu,Chakrabarty:2015foa}.  The worldsheet formalism we have developed allows us to generalize the analysis of the emission process to other supergravity modes as well as stringy quanta.

\vspace{5mm}

%%%%%%%%%%%%%%%%%%%%%%%%%%%%%%%%%%%%%%%%%%%%
%%%%%%%%%%%%%%%%%%%%%%%%%%%%%%%%%%%%%%%%%%%%
\section*{Acknowledgements}

For valuable discussions, we thank 
Iosif Bena, 
Rapha\"el Dulac, 
Samir Mathur, 
Masaki Shigemori, 
Nick Warner, 
and
Yoav Zigdon. 
The work of EJM was supported in part by DOE grant DE-SC0009924.
The work of DT was supported by a Royal Society Tata University Research Fellowship. 
For hospitality during
the course of this work, we thank the Centro de
Ciencias de Benasque and IPhT, CEA Saclay.

\vspace{5mm}

%\newpage

\begin{appendix}

%%%%%%%%%%%%%%%%%%%%%%%%%%%%%%%%%
\section{The scaling limit described by worldsheet models}
\label{sec:asymflat}

In this appendix we discuss the normalization of the dilaton and the conditions on moduli and quantum numbers such that we work consistently in the NS5-F1 frame with a local string coupling that is everywhere small.
We first consider the simple geometry where the sources are all at $r_\flat=0$.  The geometry is
\begin{align}
ds^2 &= \frac{1}{H_1}\big(-dt_\flat^2+dy_\flat^2\big)+H_5\big(dr_\flat^2+r_\flat^2d\Omega_3^2\big) + d\zz\cdot d\zz ~,
\\[.2cm]
e^{2\Phi} &= g_s^2\frac{H_5}{H_1}
~~,~~~~
H_{1,5} = 1+\frac{Q_{1,5}}{r_\flat^2}~,
\end{align}
where 
\be
\label{Qdef}
Q_5 = n_5\alpha'
~~,~~~~
Q_1 = \frac{g_s^2 n_1 (\alpha')^3}{V_4} ~.
\ee
Henceforth we work in units in which $\alpha'=1$.
Now let us rescale $r_\flat=g_s \tilde r$, so that
\begin{align}
ds^2 &= \frac{\tilde r^2}{\tilde r^2+n_1/V_4}\big(-dt_\flat^2+dy_\flat^2\big) + \Big( g_s^2 + \frac{n_5}{\tilde r^2}\Big)\big(d\tilde r^2+\tilde r^2d\Omega_3^2\big) + d\zz\cdot d\zz 
\nn\\[.2cm]
e^{2\Phi} &= \frac{g_s^2\tilde r^2+n_5}{\tilde r^2 + n_1/V_4} ~.
\end{align}
The dilaton evolves from its fixed scalar value $n_5 V_4/n_1$ at $\tilde r=0$ to its asymptotic value of $g_s$ at $\tilde r\to\infty$. In the intermediate range of $\tilde{r}$, that is when  $\tilde r \gg \sqrt{n_1/V_4}\equiv r_1$ and $\tilde r \ll \sqrt{n_5/g_s^2}\equiv r_5$, one has a region of linear dilaton. To be consistently in the NS5-F1 frame, we thus arrange 
\be \label{eq:small-dil-conds}
g_s \,\ll\, 1 \,, 
\qquad
\frac{n_5}{n_1}V_4 \,\ll\, 1 \,, 
\ee
as well as $V_4 \ge 1$, so that the string coupling in both 10D and 6D is small in both asymptotic and small $\tilde{r}$ regions. The second of the above two conditions should be thought of as working at parametrically large $n_1$.

The proper size of $\bS^1_y$ grows from zero at $\tilde r=0$ to $R_y$ at $\tilde r=r_1$, at which point it saturates and stays that size all the way out to $\tilde r\to\infty$.  The fivebrane decoupling limit is, as usual, $g_s\to 0$, which sends $r_5\to\infty$ and then the linear dilaton regime extends all the way out to infinite $\tilde r$.

The geometry is not yet in $AdS_3$ form in the small $\tilde r$ region; we thus do a further rescaling 
\be
\label{adef}
r_\flat = g_s\tilde r = a r
~~,~~~~
a^2 \equiv {\frac{g_s^2n_1n_5}{V_4 R_y^2}}
~~,~~~~
t=t_\flat/R_y
~~,~~~~
y=y_\flat/R_y
~, 
\ee
to arrive at
\begin{align}
ds^2 &= \frac{r^2R_y^2}{r^2+R_y^2/n_5}\big(-dt^2+dy^2\big) + \Big(\frac{g_s^2n_1n_5}{V_4R_y^2} + \frac{n_5}{r^2}\Big)\big(dr^2+r^2\Omega_3^2\big) + d\zz\cdot d\zz ~,
\nn\\[.2cm]
e^{2\Phi} &= \frac{g_s^2r^2 + V_4R_y^2/n_1}{r^2 + R_y^2/n_5}  ~.
\end{align}
Thus, in the region $r\ll R_y\sqrt{n_5}$, we have a canonical $AdS_3$ metric with radius $n_5$ in string units.  The fivebrane decoupling limit, 
$g_s \to 0$ at fixed $r$, has a linear dilaton regime which extends over $R_y/\sqrt{n_5} \ll r <\infty$.

If we want a decoupling limit in which $AdS_3$ fills all of spacetime, we simply scale $R_y\to\infty$ in these coordinates; the constant terms in both harmonic functions go away.  Note that, in the original flat coordinates, we don't have to scale $g_s$ in any particular way in order to achieve this result.  In that case, we are holding the asymptotic region fixed and blowing up the $y$ circle while sending the energy to zero like $1/R_y$ to obtain a decoupled theory with nontrivial dynamics.
Finally, note that when $r_1=r_5$, \ie\ $g_s^2 n_1=n_5 V_4$, the dilaton is constant in the above solution. Indeed, the geometry interpolates directly from $AdS_3$ to asymptotically flat spacetime at the scale $r=R_y$, 
without an intervening linear dilaton region,

When we take the fivebrane decoupling limit, we rescale the radial coordinate with $g_s$ as above. Correspondingly, we define $\medtilde{Q}_1$ via $Q_1 = g_s^2 \medtilde{Q}_1$ and hold $\medtilde{Q}_1$ finite as we scale $g_s \to 0$. Then, in units in which $\alpha'=1$, and ignoring factors of $2\pi$, we have
\be
\medtilde{Q}_1  \,=\, \frac{n_1}{V_4} \,, \qquad
Q_5 \,=\, n_5 \,.
\ee
Then, the condition that the string coupling is weak at small $\tilde{r}$ can be expressed as
\be
\label{eq:n1large}
n_1 \,\gg\, n_5 V_4  \qquad
\Rightarrow
\qquad
\medtilde{Q}_1 \,\gg\, Q_5 \,.  
\ee
We emphasize that the original $Q_1$ has been scaled towards zero, and is therefore parametrically smaller than $Q_5$. However, $n_1$ is parametrically large, and has scaled out of the metric and B-field; it only appears in the dilaton. 

For the GLMT backgrounds, the value of $\Sigma$ at $\rho=0$ depends on $\theta$, and ranges between the values
\begin{equation}
\label{eq:dil-cap}
(s+1)^2+\frac{\sfk^2 R_y^2}{n_5}
\qquad 
\mathrm{and}
\qquad 
s^2+\frac{\sfk^2 R_y^2}{n_5} ~.
\end{equation}
For large $R_y$, to leading order these are both 
$\sfk^2 R_y^2/n_5$. So at large $R_y$, the value of the dilaton in the cap is again given to leading order by
\begin{equation}
\label{eq:dil-app}
e^{2\Phi} \,=\, \frac{n_5 V_4}{n_1} \;,
\end{equation}
where the first subleading corrections are down by a factor of $n_5/R_y^2$. Thus, the condition \eqref{eq:n1large} ensures that we have a parametrically small dilaton everywhere in the backgrounds that we study in this work.

\end{appendix}

%%%%%%%%%%%%%%%%%%%%%%%%%%%%%%%%%%%%%%
%%%%%%%%%%%%%%%%%%%%%%%%%%%%%%%%%%%%%%

\newpage
%\vskip 2cm

\bibliographystyle{JHEP}      

\bibliography{microstates}

\providecommand{\href}[2]{#2}\begingroup\raggedright\begin{thebibliography}{10}

\bibitem{Lunin:2001fv}
O.~Lunin and S.~D. Mathur, \emph{{Metric of the multiply wound rotating
  string}}, \href{http://dx.doi.org/10.1016/S0550-3213(01)00321-2}{\emph{Nucl.
  Phys.} {\bfseries B610} (2001) 49--76},
  [\href{https://arxiv.org/abs/hep-th/0105136}{{\ttfamily hep-th/0105136}}].

\bibitem{Lunin:2001jy}
O.~Lunin and S.~D. Mathur, \emph{{AdS/CFT duality and the black hole
  information paradox}},
  \href{http://dx.doi.org/10.1016/S0550-3213(01)00620-4}{\emph{Nucl. Phys.}
  {\bfseries B623} (2002) 342--394},
  [\href{https://arxiv.org/abs/hep-th/0109154}{{\ttfamily hep-th/0109154}}].

\bibitem{Rychkov:2005ji}
V.~S. Rychkov, \emph{{D1-D5 black hole microstate counting from supergravity}},
  \href{http://dx.doi.org/10.1088/1126-6708/2006/01/063}{\emph{JHEP} {\bfseries
  01} (2006) 063}, [\href{https://arxiv.org/abs/hep-th/0512053}{{\ttfamily
  hep-th/0512053}}].

\bibitem{Kanitscheider:2007wq}
I.~Kanitscheider, K.~Skenderis and M.~Taylor, \emph{{Fuzzballs with internal
  excitations}}, {\emph{JHEP} {\bfseries 06} (2007) 056},
  [\href{https://arxiv.org/abs/0704.0690}{{\ttfamily 0704.0690}}].

\bibitem{Giusto:2015dfa}
S.~Giusto, E.~Moscato and R.~Russo, \emph{{AdS$_{3}$ holography for 1/4 and 1/8
  BPS geometries}},
  \href{http://dx.doi.org/10.1007/JHEP11(2015)004}{\emph{JHEP} {\bfseries 11}
  (2015) 004}, [\href{https://arxiv.org/abs/1507.00945}{{\ttfamily
  1507.00945}}].

\bibitem{Balasubramanian:2000rt}
V.~Balasubramanian, J.~de~Boer, E.~Keski-Vakkuri and S.~F. Ross,
  \emph{{Supersymmetric conical defects: Towards a string theoretic description
  of black hole formation}},
  \href{http://dx.doi.org/10.1103/PhysRevD.64.064011}{\emph{Phys. Rev.}
  {\bfseries D64} (2001) 064011},
  [\href{https://arxiv.org/abs/hep-th/0011217}{{\ttfamily hep-th/0011217}}].

\bibitem{Maldacena:2000dr}
J.~M. Maldacena and L.~Maoz, \emph{{De-singularization by rotation}},
  {\emph{JHEP} {\bfseries 12} (2002) 055},
  [\href{https://arxiv.org/abs/hep-th/0012025}{{\ttfamily hep-th/0012025}}].

\bibitem{Lunin:2004uu}
O.~Lunin, \emph{{Adding momentum to D1-D5 system}},
  \href{http://dx.doi.org/10.1088/1126-6708/2004/04/054}{\emph{JHEP} {\bfseries
  04} (2004) 054}, [\href{https://arxiv.org/abs/hep-th/0404006}{{\ttfamily
  hep-th/0404006}}].

\bibitem{Giusto:2004id}
S.~Giusto, S.~D. Mathur and A.~Saxena, \emph{{Dual geometries for a set of
  3-charge microstates}},
  \href{http://dx.doi.org/10.1016/j.nuclphysb.2004.09.001}{\emph{Nucl. Phys.}
  {\bfseries B701} (2004) 357--379},
  [\href{https://arxiv.org/abs/hep-th/0405017}{{\ttfamily hep-th/0405017}}].

\bibitem{Giusto:2004ip}
S.~Giusto, S.~D. Mathur and A.~Saxena, \emph{{3-charge geometries and their CFT
  duals}}, \href{http://dx.doi.org/10.1016/j.nuclphysb.2005.01.009}{\emph{Nucl.
  Phys.} {\bfseries B710} (2005) 425--463},
  [\href{https://arxiv.org/abs/hep-th/0406103}{{\ttfamily hep-th/0406103}}].

\bibitem{Jejjala:2005yu}
V.~Jejjala, O.~Madden, S.~F. Ross and G.~Titchener, \emph{{Non-supersymmetric
  smooth geometries and D1-D5-P bound states}},
  \href{http://dx.doi.org/10.1103/PhysRevD.71.124030}{\emph{Phys. Rev.}
  {\bfseries D71} (2005) 124030},
  [\href{https://arxiv.org/abs/hep-th/0504181}{{\ttfamily hep-th/0504181}}].

\bibitem{Giusto:2012yz}
S.~Giusto, O.~Lunin, S.~D. Mathur and D.~Turton, \emph{{D1-D5-P microstates at
  the cap}}, \href{http://dx.doi.org/10.1007/JHEP02(2013)050}{\emph{JHEP}
  {\bfseries 1302} (2013) 050},
  [\href{https://arxiv.org/abs/1211.0306}{{\ttfamily 1211.0306}}].

\bibitem{Chakrabarty:2015foa}
B.~Chakrabarty, D.~Turton and A.~Virmani, \emph{{Holographic description of
  non-supersymmetric orbifolded D1-D5-P solutions}},
  \href{http://dx.doi.org/10.1007/JHEP11(2015)063}{\emph{JHEP} {\bfseries 11}
  (2015) 063}, [\href{https://arxiv.org/abs/1508.01231}{{\ttfamily
  1508.01231}}].

\bibitem{Martinec:2017ztd}
E.~J. Martinec and S.~Massai, \emph{{String Theory of Supertubes}},
  \href{http://dx.doi.org/10.1007/JHEP07(2018)163}{\emph{JHEP} {\bfseries 07}
  (2018) 163}, [\href{https://arxiv.org/abs/1705.10844}{{\ttfamily
  1705.10844}}].

\bibitem{Bufalini:2021ndn}
D.~Bufalini, S.~Iguri, N.~Kovensky and D.~Turton, \emph{{Black hole microstates
  from the worldsheet}},
  \href{http://dx.doi.org/10.1007/JHEP08(2021)011}{\emph{JHEP} {\bfseries 08}
  (2021) 011}, [\href{https://arxiv.org/abs/2105.02255}{{\ttfamily
  2105.02255}}].

\bibitem{Skenderis:2008qn}
K.~Skenderis and M.~Taylor, \emph{{The fuzzball proposal for black holes}},
  \href{http://dx.doi.org/10.1016/j.physrep.2008.08.001}{\emph{Phys. Rept.}
  {\bfseries 467} (2008) 117--171},
  [\href{https://arxiv.org/abs/0804.0552}{{\ttfamily 0804.0552}}].

\bibitem{Galliani:2016cai}
A.~Galliani, S.~Giusto, E.~Moscato and R.~Russo, \emph{{Correlators at large c
  without information loss}},
  \href{http://dx.doi.org/10.1007/JHEP09(2016)065}{\emph{JHEP} {\bfseries 09}
  (2016) 065}, [\href{https://arxiv.org/abs/1606.01119}{{\ttfamily
  1606.01119}}].

\bibitem{Chakrabarty:2021sff}
B.~Chakrabarty, S.~Rawash and D.~Turton, \emph{{Shockwaves in black hole
  microstate geometries}},
  \href{http://dx.doi.org/10.1007/JHEP02(2022)202}{\emph{JHEP} {\bfseries 02}
  (2022) 202}, [\href{https://arxiv.org/abs/2112.08378}{{\ttfamily
  2112.08378}}].

\bibitem{Martinec:2018nco}
E.~J. Martinec, S.~Massai and D.~Turton, \emph{{String dynamics in NS5-F1-P
  geometries}}, \href{http://dx.doi.org/10.1007/JHEP09(2018)031}{\emph{JHEP}
  {\bfseries 09} (2018) 031},
  [\href{https://arxiv.org/abs/1803.08505}{{\ttfamily 1803.08505}}].

\bibitem{Martinec:2019wzw}
E.~J. Martinec, S.~Massai and D.~Turton, \emph{{Little Strings, Long Strings,
  and Fuzzballs}}, \href{http://dx.doi.org/10.1007/JHEP11(2019)019}{\emph{JHEP}
  {\bfseries 11} (2019) 019},
  [\href{https://arxiv.org/abs/1906.11473}{{\ttfamily 1906.11473}}].

\bibitem{Martinec:2020gkv}
E.~J. Martinec, S.~Massai and D.~Turton, \emph{{Stringy Structure at the BPS
  Bound}}, \href{http://dx.doi.org/10.1007/JHEP12(2020)135}{\emph{JHEP}
  {\bfseries 12} (2020) 135},
  [\href{https://arxiv.org/abs/2005.12344}{{\ttfamily 2005.12344}}].

\bibitem{Bufalini:2022wyp}
D.~Bufalini, S.~Iguri, N.~Kovensky and D.~Turton, \emph{{Worldsheet Correlators
  in Black Hole Microstates}},
  \href{http://dx.doi.org/10.1103/PhysRevLett.129.121603}{\emph{Phys. Rev.
  Lett.} {\bfseries 129} (2022) 121603},
  [\href{https://arxiv.org/abs/2203.13828}{{\ttfamily 2203.13828}}].

\bibitem{Bufalini:2022wzu}
D.~Bufalini, S.~Iguri, N.~Kovensky and D.~Turton, \emph{{Worldsheet computation
  of heavy-light correlators}},
  \href{http://dx.doi.org/10.1007/JHEP03(2023)066}{\emph{JHEP} {\bfseries 03}
  (2023) 066}, [\href{https://arxiv.org/abs/2210.15313}{{\ttfamily
  2210.15313}}].

\bibitem{Martinec:2022okx}
E.~J. Martinec, S.~Massai and D.~Turton, \emph{{On the BPS Sector in AdS3/CFT2
  Holography}}, \href{http://dx.doi.org/10.1002/prop.202300015}{\emph{Fortsch.
  Phys.} {\bfseries 71} (2023) 2300015},
  [\href{https://arxiv.org/abs/2211.12476}{{\ttfamily 2211.12476}}].

\bibitem{Eperon:2016cdd}
F.~C. Eperon, H.~S. Reall and J.~E. Santos, \emph{{Instability of
  supersymmetric microstate geometries}},
  \href{http://dx.doi.org/10.1007/JHEP10(2016)031}{\emph{JHEP} {\bfseries 10}
  (2016) 031}, [\href{https://arxiv.org/abs/1607.06828}{{\ttfamily
  1607.06828}}].

\bibitem{Marolf:2016nwu}
D.~Marolf, B.~Michel and A.~Puhm, \emph{{A rough end for smooth microstate
  geometries}}, \href{http://dx.doi.org/10.1007/JHEP05(2017)021}{\emph{JHEP}
  {\bfseries 05} (2017) 021},
  [\href{https://arxiv.org/abs/1612.05235}{{\ttfamily 1612.05235}}].

\bibitem{Bena:2018mpb}
I.~Bena, E.~J. Martinec, R.~Walker and N.~P. Warner, \emph{{Early Scrambling
  and Capped BTZ Geometries}},
  \href{http://dx.doi.org/10.1007/JHEP04(2019)126}{\emph{JHEP} {\bfseries 04}
  (2019) 126}, [\href{https://arxiv.org/abs/1812.05110}{{\ttfamily
  1812.05110}}].

\bibitem{Friedman:1978ygc}
J.~L. Friedman, \emph{{Ergosphere instability}},
  \href{http://dx.doi.org/10.1007/BF01196933}{\emph{Commun. Math. Phys.}
  {\bfseries 63} (1978) 243--255}.

\bibitem{Cardoso:2005gj}
V.~Cardoso, O.~J.~C. Dias, J.~L. Hovdebo and R.~C. Myers, \emph{{Instability of
  non-supersymmetric smooth geometries}},
  \href{http://dx.doi.org/10.1103/PhysRevD.73.064031}{\emph{Phys. Rev.}
  {\bfseries D73} (2006) 064031},
  [\href{https://arxiv.org/abs/hep-th/0512277}{{\ttfamily hep-th/0512277}}].

\bibitem{Chowdhury:2007jx}
B.~D. Chowdhury and S.~D. Mathur, \emph{{Radiation from the non-extremal
  fuzzball}},
  \href{http://dx.doi.org/10.1088/0264-9381/25/13/135005}{\emph{Class. Quant.
  Grav.} {\bfseries 25} (2008) 135005},
  [\href{https://arxiv.org/abs/0711.4817}{{\ttfamily 0711.4817}}].

\bibitem{Chowdhury:2008bd}
B.~D. Chowdhury and S.~D. Mathur, \emph{{Pair creation in non-extremal fuzzball
  geometries}},
  \href{http://dx.doi.org/10.1088/0264-9381/25/22/225021}{\emph{Class. Quant.
  Grav.} {\bfseries 25} (2008) 225021},
  [\href{https://arxiv.org/abs/0806.2309}{{\ttfamily 0806.2309}}].

\bibitem{Chowdhury:2008uj}
B.~D. Chowdhury and S.~D. Mathur, \emph{{Non-extremal fuzzballs and ergoregion
  emission}},
  \href{http://dx.doi.org/10.1088/0264-9381/26/3/035006}{\emph{Class. Quant.
  Grav.} {\bfseries 26} (2009) 035006},
  [\href{https://arxiv.org/abs/0810.2951}{{\ttfamily 0810.2951}}].

\bibitem{Avery:2009tu}
S.~G. Avery, B.~D. Chowdhury and S.~D. Mathur, \emph{{Emission from the D1D5
  CFT}}, \href{http://dx.doi.org/10.1088/1126-6708/2009/10/065}{\emph{JHEP}
  {\bfseries 10} (2009) 065},
  [\href{https://arxiv.org/abs/0906.2015}{{\ttfamily 0906.2015}}].

\bibitem{Chervonyi:2013eja}
Y.~Chervonyi and O.~Lunin, \emph{{(Non)-Integrability of Geodesics in D-brane
  Backgrounds}}, \href{http://dx.doi.org/10.1007/JHEP02(2014)061}{\emph{JHEP}
  {\bfseries 02} (2014) 061},
  [\href{https://arxiv.org/abs/1311.1521}{{\ttfamily 1311.1521}}].

\bibitem{Bianchi:2017sds}
M.~Bianchi, D.~Consoli and J.~F. Morales, \emph{{Probing Fuzzballs with
  Particles, Waves and Strings}},
  \href{http://dx.doi.org/10.1007/JHEP06(2018)157}{\emph{JHEP} {\bfseries 06}
  (2018) 157}, [\href{https://arxiv.org/abs/1711.10287}{{\ttfamily
  1711.10287}}].

\bibitem{Gutowski:2003rg}
J.~B. Gutowski, D.~Martelli and H.~S. Reall, \emph{{All supersymmetric
  solutions of minimal supergravity in six dimensions}},
  \href{http://dx.doi.org/10.1088/0264-9381/20/23/008}{\emph{Class. Quant.
  Grav.} {\bfseries 20} (2003) 5049--5078},
  [\href{https://arxiv.org/abs/hep-th/0306235}{{\ttfamily hep-th/0306235}}].

\bibitem{Pelavas:2005}
N.~Pelavas, \emph{{Timelike Killing vectors and ergo surfaces in
  non-asymptotically flat spacetimes}}, {\emph{Gen. Rel. Grav.} {\bfseries 37}
  (2005) 313}.

\bibitem{Gibbons:2013tqa}
G.~Gibbons and N.~Warner, \emph{{Global structure of five-dimensional
  fuzzballs}},
  \href{http://dx.doi.org/10.1088/0264-9381/31/2/025016}{\emph{Class.Quant.Grav.}
  {\bfseries 31} (2014) 025016},
  [\href{https://arxiv.org/abs/1305.0957}{{\ttfamily 1305.0957}}].

\bibitem{Chakrabarty:2019ujg}
B.~Chakrabarty, D.~Ghosh and A.~Virmani, \emph{{Quasinormal modes of
  supersymmetric microstate geometries from the D1-D5 CFT}},
  \href{http://dx.doi.org/10.1007/JHEP10(2019)072}{\emph{JHEP} {\bfseries 10}
  (2019) 072}, [\href{https://arxiv.org/abs/1908.01461}{{\ttfamily
  1908.01461}}].

\bibitem{Maldacena:2000hw}
J.~M. Maldacena and H.~Ooguri, \emph{{Strings in AdS(3) and SL(2,R) WZW model
  1.: The Spectrum}}, \href{http://dx.doi.org/10.1063/1.1377273}{\emph{J. Math.
  Phys.} {\bfseries 42} (2001) 2929--2960},
  [\href{https://arxiv.org/abs/hep-th/0001053}{{\ttfamily hep-th/0001053}}].

\bibitem{Giveon:1998ns}
A.~Giveon, D.~Kutasov and N.~Seiberg, \emph{{Comments on string theory on
  AdS(3)}}, {\emph{Adv. Theor. Math. Phys.} {\bfseries 2} (1998) 733--780},
  [\href{https://arxiv.org/abs/hep-th/9806194}{{\ttfamily hep-th/9806194}}].

\bibitem{Argurio:2000tb}
R.~Argurio, A.~Giveon and A.~Shomer, \emph{{Superstrings on AdS(3) and
  symmetric products}},
  \href{http://dx.doi.org/10.1088/1126-6708/2000/12/003}{\emph{JHEP} {\bfseries
  12} (2000) 003}, [\href{https://arxiv.org/abs/hep-th/0009242}{{\ttfamily
  hep-th/0009242}}].

\bibitem{Bena:2015bea}
I.~Bena, S.~Giusto, R.~Russo, M.~Shigemori and N.~P. Warner, \emph{{Habemus
  Superstratum! A constructive proof of the existence of superstrata}},
  \href{http://dx.doi.org/10.1007/JHEP05(2015)110}{\emph{JHEP} {\bfseries 05}
  (2015) 110}, [\href{https://arxiv.org/abs/1503.01463}{{\ttfamily
  1503.01463}}].

\bibitem{Bena:2016agb}
I.~Bena, E.~Martinec, D.~Turton and N.~P. Warner, \emph{{Momentum Fractionation
  on Superstrata}},
  \href{http://dx.doi.org/10.1007/JHEP05(2016)064}{\emph{JHEP} {\bfseries 05}
  (2016) 064}, [\href{https://arxiv.org/abs/1601.05805}{{\ttfamily
  1601.05805}}].

\bibitem{Bena:2016ypk}
I.~Bena, S.~Giusto, E.~J. Martinec, R.~Russo, M.~Shigemori, D.~Turton et~al.,
  \emph{{Smooth horizonless geometries deep inside the black-hole regime}},
  \href{http://dx.doi.org/10.1103/PhysRevLett.117.201601}{\emph{Phys. Rev.
  Lett.} {\bfseries 117} (2016) 201601},
  [\href{https://arxiv.org/abs/1607.03908}{{\ttfamily 1607.03908}}].

\bibitem{Bena:2017xbt}
I.~Bena, S.~Giusto, E.~J. Martinec, R.~Russo, M.~Shigemori, D.~Turton et~al.,
  \emph{{Asymptotically-flat supergravity solutions deep inside the black-hole
  regime}}, \href{http://dx.doi.org/10.1007/JHEP02(2018)014}{\emph{JHEP}
  {\bfseries 02} (2018) 014},
  [\href{https://arxiv.org/abs/1711.10474}{{\ttfamily 1711.10474}}].

\bibitem{Bena:2018bbd}
I.~Bena, P.~Heidmann and D.~Turton, \emph{{AdS$_{2}$ holography: mind the
  cap}}, \href{http://dx.doi.org/10.1007/JHEP12(2018)028}{\emph{JHEP}
  {\bfseries 12} (2018) 028},
  [\href{https://arxiv.org/abs/1806.02834}{{\ttfamily 1806.02834}}].

\bibitem{Ceplak:2018pws}
N.~\v{C}eplak, R.~Russo and M.~Shigemori, \emph{{Supercharging Superstrata}},
  \href{http://dx.doi.org/10.1007/JHEP03(2019)095}{\emph{JHEP} {\bfseries 03}
  (2019) 095}, [\href{https://arxiv.org/abs/1812.08761}{{\ttfamily
  1812.08761}}].

\bibitem{Heidmann:2019zws}
P.~Heidmann and N.~P. Warner, \emph{{Superstratum Symbiosis}},
  \href{http://dx.doi.org/10.1007/JHEP09(2019)059}{\emph{JHEP} {\bfseries 09}
  (2019) 059}, [\href{https://arxiv.org/abs/1903.07631}{{\ttfamily
  1903.07631}}].

\bibitem{Ganchev:2022exf}
B.~Ganchev, A.~Houppe and N.~P. Warner, \emph{{Elliptical and purely NS
  superstrata}}, \href{http://dx.doi.org/10.1007/JHEP09(2022)067}{\emph{JHEP}
  {\bfseries 09} (2022) 067},
  [\href{https://arxiv.org/abs/2207.04060}{{\ttfamily 2207.04060}}].

\bibitem{Ceplak:2022pep}
N.~\v{C}eplak, \emph{{Vector Superstrata}},
  \href{http://dx.doi.org/10.1007/JHEP08(2023)047}{\emph{JHEP} {\bfseries 08}
  (2023) 047}, [\href{https://arxiv.org/abs/2212.06947}{{\ttfamily
  2212.06947}}].

\bibitem{Ceplak:2024dbj}
N.~{\v{C}}eplak and S.~D. Hampton, \emph{{Vector superstrata. Part II}},
  \href{http://dx.doi.org/10.1007/JHEP10(2024)011}{\emph{JHEP} {\bfseries 10}
  (2024) 011}, [\href{https://arxiv.org/abs/2405.05341}{{\ttfamily
  2405.05341}}].

\bibitem{Martinec:2023zha}
E.~J. Martinec, \emph{{AdS$_{3}$ orbifolds, BTZ black holes, and holography}},
  \href{http://dx.doi.org/10.1007/JHEP10(2023)016}{\emph{JHEP} {\bfseries 10}
  (2023) 016}, [\href{https://arxiv.org/abs/2307.02559}{{\ttfamily
  2307.02559}}].

\bibitem{Avery:2010qw}
S.~G. Avery, \emph{{Using the D1D5 CFT to Understand Black Holes}},
  \href{https://arxiv.org/abs/1012.0072}{{\ttfamily 1012.0072}}.

\bibitem{Shigemori:2022gxf}
M.~Shigemori, \emph{{Superstrata on orbifolded backgrounds}},
  \href{http://dx.doi.org/10.1007/JHEP02(2023)099}{\emph{JHEP} {\bfseries 02}
  (2023) 099}, [\href{https://arxiv.org/abs/2212.13388}{{\ttfamily
  2212.13388}}].

\bibitem{Polchinski:1998rq}
J.~Polchinski, \emph{{String theory. Vol. 1: An introduction to the bosonic
  string}}.
\newblock Cambridge Monographs on Mathematical Physics. Cambridge University
  Press, 12, 2007,
  \href{http://dx.doi.org/10.1017/CBO9780511816079}{10.1017/CBO9780511816079}.

\bibitem{Balthazar:2021xeh}
B.~Balthazar, A.~Giveon, D.~Kutasov and E.~J. Martinec, \emph{{Asymptotically
  free AdS$_{3}$/CFT$_{2}$}},
  \href{http://dx.doi.org/10.1007/JHEP01(2022)008}{\emph{JHEP} {\bfseries 01}
  (2022) 008}, [\href{https://arxiv.org/abs/2109.00065}{{\ttfamily
  2109.00065}}].

\bibitem{Giusto:2019qig}
S.~Giusto, S.~Rawash and D.~Turton, \emph{{AdS$_{3}$ holography at dimension
  two}}, \href{http://dx.doi.org/10.1007/JHEP07(2019)171}{\emph{JHEP}
  {\bfseries 07} (2019) 171},
  [\href{https://arxiv.org/abs/1904.12880}{{\ttfamily 1904.12880}}].

\bibitem{Rawash:2021pik}
S.~Rawash and D.~Turton, \emph{{Supercharged AdS$_{3}$ Holography}},
  \href{http://dx.doi.org/10.1007/JHEP07(2021)178}{\emph{JHEP} {\bfseries 07}
  (2021) 178}, [\href{https://arxiv.org/abs/2105.13046}{{\ttfamily
  2105.13046}}].

\bibitem{Shigemori:2020yuo}
M.~Shigemori, \emph{{Superstrata}},
  \href{http://dx.doi.org/10.1007/s10714-020-02698-8}{\emph{Gen. Rel. Grav.}
  {\bfseries 52} (2020) 51},
  [\href{https://arxiv.org/abs/2002.01592}{{\ttfamily 2002.01592}}].

\bibitem{Bena:2011zw}
I.~Bena, B.~D. Chowdhury, J.~de~Boer, S.~El-Showk and M.~Shigemori,
  \emph{{Moulting Black Holes}},
  \href{http://dx.doi.org/10.1007/JHEP03(2012)094}{\emph{JHEP} {\bfseries 1203}
  (2012) 094}, [\href{https://arxiv.org/abs/1108.0411}{{\ttfamily 1108.0411}}].

\bibitem{Larsen:2025jqo}
F.~Larsen and S.~Lee, \emph{{Instability of Black Holes in AdS$_3 \times
  S^3$}},  \href{https://arxiv.org/abs/2507.04131}{{\ttfamily 2507.04131}}.

\bibitem{Mathur:2008kg}
S.~D. Mathur, \emph{{Tunneling into fuzzball states}},
  \href{http://dx.doi.org/10.1007/s10714-009-0837-3}{\emph{Gen. Rel. Grav.}
  {\bfseries 42} (2010) 113--118},
  [\href{https://arxiv.org/abs/0805.3716}{{\ttfamily 0805.3716}}].

\bibitem{Kraus:2015zda}
P.~Kraus and S.~D. Mathur, \emph{{Nature abhors a horizon}},
  \href{http://dx.doi.org/10.1142/S0218271815430038}{\emph{Int. J. Mod. Phys.
  D} {\bfseries 24} (2015) 1543003},
  [\href{https://arxiv.org/abs/1505.05078}{{\ttfamily 1505.05078}}].

\bibitem{BDMSTW}
I.~Bena, R.~Dulac, E.~J. Martinec, M.~Shigemori, D.~Turton and N.~P. Warner,
  \emph{To appear}, .

\bibitem{MMT-6}
E.~J. Martinec, S.~Massai and D.~Turton, \emph{To appear}, .

\end{thebibliography}\endgroup

%%%%%%%%%%%%%%%%%%%%%%%%%%%%%%%%%%%%%%
%%%%%%%%%%%%%%%%%%%%%%%%%%%%%%%%%%%%%%

\end{document}